\documentclass[aps,pre,twocolumn,showpacs,superscriptaddress,nofootinbib]{revtex4}
\usepackage{amsfonts,amssymb,amsmath,graphicx,enumerate}
\usepackage{amsthm}
\usepackage{multirow}

\newtheorem{mydef}{Definition}

\begin{document}

\title{Analytical framework for recurrence-network analysis of time series}

\author{Jonathan F. Donges}
    \email{donges@pik-potsdam.de}
    \affiliation{Potsdam Institute for Climate Impact Research, P.O. Box 601203, 14412 Potsdam, Germany}
    \affiliation{Department of Physics, Humboldt University Berlin, Newtonstr.~15, 12489 Berlin, Germany}
\author{Jobst Heitzig}
    \affiliation{Potsdam Institute for Climate Impact Research, P.O. Box 601203, 14412 Potsdam, Germany}
\author{Reik V. Donner}
    \affiliation{Potsdam Institute for Climate Impact Research, P.O. Box 601203, 14412 Potsdam, Germany}
\author{J\"urgen Kurths}
   \affiliation{Potsdam Institute for Climate Impact Research, P.O. Box 601203, 14412 Potsdam, Germany}
   \affiliation{Department of Physics, Humboldt University Berlin, Newtonstr.~15, 12489 Berlin, Germany}
   \affiliation{Institute for Complex Systems and Mathematical Biology, University of Aberdeen, Aberdeen AB24 3FX, United Kingdom}

\date{\today}

\begin{abstract}
Recurrence networks are a powerful nonlinear tool for time series analysis of complex dynamical systems. {While there are already many successful applications ranging from medicine to paleoclimatology, a solid theoretical foundation of the method has still been missing so far. Here, we interpret an $\varepsilon$-recurrence network as a discrete subnetwork of a ``continuous'' graph with uncountably many vertices and edges corresponding to the system's attractor. This step allows us to show that various statistical measures commonly used in complex network analysis can be seen as discrete estimators of newly defined continuous measures of certain complex geometric properties of the attractor on the scale given by $\varepsilon$.} In particular, we introduce local measures such as the $\varepsilon$-clustering coefficient, mesoscopic measures such as $\varepsilon$-motif density, path-based measures such as $\varepsilon$-betweennesses, and global measures such as $\varepsilon$-efficiency. This new analytical basis for the so far heuristically motivated network measures also provides an objective criterion for the choice of $\varepsilon$ via a percolation threshold, and it shows that estimation can be improved by so-called node splitting invariant versions of the measures. We finally illustrate the framework for a number of archetypical chaotic attractors such as those of the Bernoulli and logistic maps, periodic and two-dimensional quasi-periodic motions, and for hyperballs and hypercubes, by deriving analytical expressions for the novel measures and comparing them with data from numerical experiments. More generally, the theoretical framework put forward in this work describes random geometric graphs and other networks with spatial constraints which appear frequently in disciplines ranging from biology to climate science.
\end{abstract}

\maketitle

\section{Introduction}

{Analogies are a fundamental motor of innovation in physics and other disciplines, since they allow the transfer of theoretical insights, results and techniques from one field to the other. In the last years, complex network theory has been particularly successful in providing unifying concepts and methods for understanding the structure and dynamics of complex systems in many areas of science, ranging from power grids over social networks to neuronal networks~\cite{Newman2003,Boccaletti2006,Cohenbook2010,Newmanbook2010}. Similarly, nonlinear time series analysis aims to gain insights on a wide variety of natural, technological, and experimental dynamical systems drawing on a generic body of theory and methods~\cite{Kantz1997}. 

By exploiting analogies in the structure and description of complex networks and dynamical systems, a number of new network-based techniques for nonlinear time series analysis have been proposed recently~\cite{Donner2011a}. The first class of these methods makes use of graph representations of certain similarity relationships between state vectors or groups of state vectors (e.g., cycles) in phase space. It includes transition networks based on a coarse-graining of phase space~\cite{Nicolis2005}, cycle networks~\cite{Zhang2006}, correlation networks~\cite{Yang2008}, $k$-nearest-neighbor~\cite{Shimada2008}, and adaptive nearest neighbor networks~\cite{Xu2008,Xiang2012} as well as $\varepsilon$-recurrence networks~\cite{Marwan2009,Donner2010b}. The latter three techniques harness the fundamental analogy between the Poincar\'e recurrence structure~\cite{Poincare1890} of a time series in phase space, which is commonly represented by a binary recurrence matrix and allows us to recover basic dynamical invariants of the underlying system~\cite{Marwan2007}, and the binary adjacency matrix describing a complex network. Other methods such as visibility graphs~\cite{Lacasa2008} work in the time domain and focus on studying stochastic properties of time series. Aside from these network-based approaches for investigating (possibly multivariate) time series from isolated dynamical systems, efforts have been spent for developing techniques for studying fields of time series such as functional brain networks~\cite{Zhou2006,Zamora2009,Zamora2010} in the neurosciences or climate networks~\cite{Tsonis2004,Donges2009a,Donges2009b,Donges2011a} in climatology. In summary, all methods mentioned above propose a mapping from the time series to the network domain and then proceed to interpret the statistical properties of the resulting (usually complex) network in terms of the underlying system's dynamical properties.}

While these interpretations are mostly based on empirical findings for paradigmatic model systems and heuristic arguments, only a few rigorous results are available. So far, Lacasa \textit{et al.} have pointed out a relationship between the scaling exponent of the degree distribution $p_k(k) \propto k^{-\gamma}$ in visibility graphs constructed from fractional Brownian motion and the Hurst exponent~\cite{Lacasa2009}. Furthermore, close relationships between the transitivity properties (network transitivity and local clustering coefficients)~\cite{Donner2010dimensions} as well as the degree distribution's power-law scaling exponent $\gamma$~\cite{Zou2011} of $\varepsilon$-recurrence networks and the (fractal) global and local dimensionality of the attracting set underlying the time series have been found. Constituting random geometric graphs~\cite{Dall2002}, $\varepsilon$-recurrence networks represent the geometry induced by the time series in phase space in a simple and well-defined way. This enabled Donner \textit{et al.}~\cite{Donner2010b,Donner2010dimensions} to define continuous transitivity properties depending solely on the geometry of the (attracting) set $S$ and the probability density function $p(x)$. These can in turn be calculated analytically for paradigmatic model systems with smooth and self-similar geometry and are approximated by the corresponding discrete $\varepsilon$-recurrence network measures. {Notably, most kinds of time series networks proposed so far are spatial networks~\cite{Barthelemy2011}, since vertices are embedded either in phase space or on the time axis, implying that general results obtained for this class of networks are applicable to time series networks as well.}

Recently, $\varepsilon$-recurrence networks have been demonstrated to be a particularly useful tool in diverse applications of nonlinear time series analysis ranging from model systems~\cite{Marwan2009,Donner2010a,Donner2010b,Donner2010c,Senthilkumar2010,Donner2011a,Donner2010dimensions,Zou2010,Li2011,Strozzi2011,Zou2012} via experimental data~\cite{Gao2009a,Gao2009b,Gao2010,Marwan2010} to recent and paleo-climate records~\cite{Marwan2009,Donner2011a,Donges2011b,Donges2011c,Hirata2011} as well as financial time series~\cite{Donner2010c}. They allow us to uncover complex bifurcation scenarios~\cite{Marwan2009,Donges2011b} and to reliably distinguish between chaotic and non-chaotic dynamics~\cite{Zou2010}. Furthermore, the local and global transitivity characteristics of $\varepsilon$-recurrence networks have been shown to enable us to trace unstable periodic orbits~\cite{Donner2010b} and to define alternative notions of fractal dimension~\cite{Donner2010dimensions} independently of earlier approaches. An important advantage of nonlinear $\varepsilon$-recurrence-network-based time series analysis is that it performs well with significantly shorter time series ($\mathcal{O}(10^2)$ data points~\cite{Marwan2009,Zou2010,Donges2011b}) than required by classical techniques like estimating the maximum Lyapunov exponent from data~\cite{Abarbanel1996,Kantz1997}. This renders $\varepsilon$-recurrence networks readily applicable to the analysis of non-stationary real world data. The method has also been applied successfully to time series with irregular sampling and/or uncertain timing of observations that are commonly found in the geosciences or in astrophysics~\cite{Donges2011b,Donges2011c}.

An $\varepsilon$-recurrence network is completely defined by its adjacency matrix $A_{ij}(\varepsilon)$ which is obtained from a (multidimensional) time series $x(t_i)$, $i=1,\dots,N$, by
\begin{equation}
A_{ij}(\varepsilon) = \Theta\left(\varepsilon - \|x(t_i) - x(t_j)\| \right) - \delta_{ij}, \label{eq:rn_def}
\end{equation}
where $\Theta(\cdot)$ is the Heaviside function, $\varepsilon$ a threshold used for defining the neighborhood of a state vector $x(t_i)$, $\|\cdot\|$ some norm, and $\delta_{ij}$ denotes Kronecker's delta introduced to avoid self-loops in the network. Given univariate observational or experimental time series, it is usually necessary to reconstruct the corresponding system's trajectory in some higher dimensional phase space to recover its recurrence structure reliably (e.g., by time-delay embedding~\cite{Packard1980,Takens1981}).

Within the recurrence network, vertices represent observations or state vectors in phase space, while edges indicate a close proximity between two state vectors. Recurrence networks and their statistical properties are related, but complementary to the established concepts of recurrence plots (the recurrence matrix is given by $R_{ij}(\varepsilon) = A_{ij}(\varepsilon) + \delta_{ij}$) and recurrence quantification analysis (RQA)~\cite{Marwan2007}. In contrast to RQA, which considers temporal dependencies between observations in form of diagonal and vertical line structures in the recurrence plot, recurrence network analysis discards all temporal information and solely quantifies the geometry of the underlying set $S$ (e.g., an attractor)~\cite{Donner2010b,Donges2011b}.

Given the diverse and successful applications of $\varepsilon$-recurrence network analysis reported in the literature, it is important to establish a firm theoretical foundation for advancing the understanding of the method. Building on earlier work~\cite{Donner2010dimensions}, we propose here an analytical framework for $\varepsilon$-recurrence network analysis of time series encompassing neighborhood-based transitivity measures, mesoscopic measures relying on network motifs~\cite{Milo2002}, path-based network characteristics as well as spectral and random walk-based measures. Specifically, our theory describes all graph-theoretical recurrence network quantifiers that have been used in the literature so far~\cite{Donner2010b}. Beyond forming a solid theoretical basis for this modern nonlinear approach to time series analysis and fostering its detailed understanding in a way comparable to that of standard linear time series analysis~\cite{BrockwellDavies2002}, our analytical framework opens several avenues for practically improving the method when dealing with finite (real-world) time series: (i) We are able to obtain closed-form analytical results for paradigmatic model systems with stochastic (uniform and Gaussian noise) and deterministic (periodic, quasi-periodic, and chaotic) dynamics. These can in turn be harnessed as a benchmark for the discrete standard estimators from complex network theory which have been employed so far~\cite{Donner2010b}, e.g., for assessing the estimators' bias and variance. (ii) This bottom-up approach allows us to design improved, weighted statistical estimators~\cite{Heitzig2011} which may be more appropriate in specific situations. (iii) Moreover, our framework enables us to derive rigorous bounds for feasible values of the recurrence threshold $\varepsilon$, the most important parameter of the method, the choice of which is critical when analyzing finite (experimental) time series~\cite{Donner2010a}. {We will argue in Sec.~\ref{sec:discussion} that the concepts and measures developed in this paper can be readily generalized to describe the structure of a wider class of spatial networks, e.g., random networks with an arbitrary prescribed edge length distribution $p_l(l)$.}

This paper is organized as follows: We introduce a continuous framework for recurrence network analysis in Sec.~\ref{sec:continuous_framework}. After reviewing the corresponding discrete estimators (Sec.~\ref{sec:estimators}), we present examples ranging from periodic and quasi-periodic dynamics and higher dimensional symmetric sets over chaotic maps to stochastic processes and compare some of the results to discrete estimates (Sec.~\ref{sec:examples}). We conclude with a discussion of these achievements (Sec.~\ref{sec:discussion}).

\section{Continuous framework}
\label{sec:continuous_framework}

\subsection{General setting}

\begin{figure}[tbp]
   \centering
   \includegraphics[width=\columnwidth]{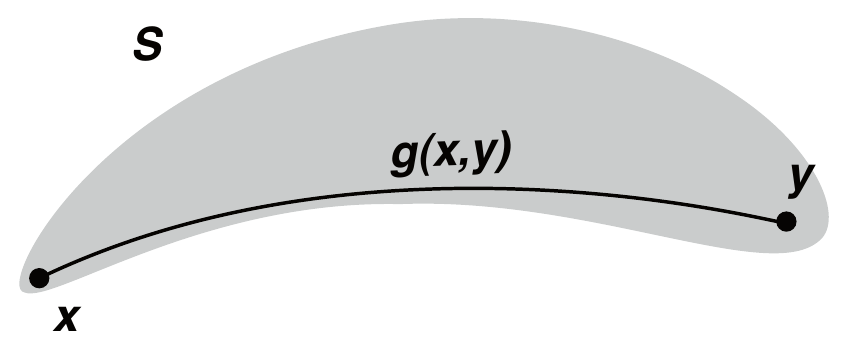}
   \caption{Illustration of a set $S$ (gray), where $g(x,y)$ denotes the geodesic distance between $x,y \in S$.}
   \label{fig:set_sketch}
\end{figure}

Let us consider a 
path-connected 
Lebesgue-measurable subset $S \subset X$ of an $m$-dimensional compact smooth manifold $X$ 
with a non-vanishing continuous probability density function $p:S\to(0,\infty)$ with $\int_S d^mx\,p(x) = 1$. 
We will use the abbreviation 
$\int d\mu(x)=\int_S d^mx\,p(x)$ 
throughout the paper, where $\mu$ is a probability measure on $S$ (Fig.~\ref{fig:set_sketch}). Then we can define 
``continuous'' 
equivalents of all relevant graph-theoretical measures for $\varepsilon$-recurrence networks which may be approximated by calculating their discrete counterparts in the limit $\varepsilon \to 0$, $N \to \infty$ (Sec.~\ref{sec:estimators}). Here $\varepsilon$ is the threshold used for network construction~(Eq.~(\ref{eq:rn_def})) and $N$ denotes the number of data points (samples, phase space vectors, $\dots$) considered. {These measures capture the properties of a ``continuous'' network with uncountably many vertices and edges which may be defined by a continuous analog of the adjacency matrix, the \emph{adjacency function}
\begin{equation}
A(x,y) = \Theta(\varepsilon - \|x-y\|) - \delta_{xy}
\end{equation}
for all $x,y \in S$.} It is important to realize that the framework introduced in this paper is not restricted to $\varepsilon$-recurrence networks alone, but may be more generally applied to describe random geometric graphs (aka spatial networks) \cite{Dall2002,Herrmann2003,Penrose2003} and other types of networks with spatial constraints~\cite{Barnett2007,Itzkovitz2005,Barthelemy2011}.

In the following we will formally define the proposed continuous recurrence network measures and discuss their properties, interrelationships and interpretations (see Table~\ref{tab:measure_summary} for an overview). Statements made for the limits $\varepsilon\to 0$ and $x\to y$ for $x,y\in S$ should be understood to hold for smooth $S$ and $p$. We do not consider them for fractal geometries explicitly.

\begin{table*}[htdp]
\caption{A summary of the continuous geometric properties of the set $S$ and its associated probability density $p$ defined in Sec. \ref{sec:continuous_framework}. See main text for formal probabilistic interpretations.}
\label{tab:measure_summary}
\begin{center}
\begin{tabular}{llll}
Class & Name & Heuristic interpretation\\
\hline
\textit{Neighborhood-based} & & \\
Local & Continuous $\varepsilon$-degree density (Eq.~(\ref{def:contrho})) & Local density \\
& Continuous local $\varepsilon$-clustering (Eq.~(\ref{eq:local_clustering})) & Local dimension~\cite{Donner2010dimensions}\\
& Continuous $\varepsilon$-matching index (Eq.~(\ref{eq:matching_index})) & Local density gradient \\
Global & Continuous $\varepsilon$-edge density (Eq.~(\ref{eq:continuous_edge_density})) & Average local density \\
& Continuous $\varepsilon$-transitivity (Eq.~(\ref{def:cont_transitivity})) & Global dimension~\cite{Donner2010dimensions}\\
& Continuous global $\varepsilon$-clustering (Eq.~(\ref{def:cont_global_clustering})) & Average local dimension \\
& Continuous $\varepsilon$-assortativity (Eq.~(\ref{def:cont_assortativity})) & Average local density gradient \\
\hline
\textit{Mesoscopic} & Continuous $\varepsilon$-motif density (Eq.~(\ref{def:cont_motif_density})) & Higher-order density structure/ \\
 &  & Density anisotropy \\
\hline
\textit{Path-based} & & \\
Local & Continuous $\varepsilon$-closeness (Eq.~(\ref{def:cont_closeness})) & \multirow{2}{*}{{\LARGE \}} Geometric centrality} \\
& Continuous $\varepsilon$-efficiency (Eq.~(\ref{def:cont_efficiency})) &  \\
& Continuous $\varepsilon$-shortest path betweenness (Eq.~(\ref{def:cont_sp_betweenness})) & \multirow{2}{*}{{\LARGE \}} Geometric bottleneckishness} \\
& Continuous $\varepsilon$-shortest path edge betweenness (Eq.~(\ref{eq:edge_betweenness})) &  \\
Global & Continuous $\varepsilon$-average path length (Eq.~(\ref{eq:apl})) & \multirow{2}{*}{{\LARGE \}} Average separation} \\
& Continuous global $\varepsilon$-efficiency (Eq.~(\ref{eq:global_efficiency})) &  \\
& $\varepsilon$-diameter (Eq.~(\ref{eq:diameter})) & Geometric diameter \\
& $\varepsilon$-radius (Eq.~(\ref{eq:radius})) & Geometric radius \\
\end{tabular}
\end{center}
\label{default}
\end{table*}

\subsection{Neighborhood-based measures}

Among other interesting properties, it has been shown recently that the local and global transitivity properties of $\varepsilon$-recurrence networks measured by the local clustering coefficient $\mathcal{C}_i$ and the global transitivity $\mathcal{T}$, respectively, are closely related to a certain notion of the fractal dimension of an underlying set $S$ and its associated probability density $p(x)$ with $x \in S$~\cite{Donner2010dimensions}. To capture this theoretically, continuous versions of both measures denoted ${\mathcal{C}}(x;\varepsilon)$ and ${\mathcal{T}}(\varepsilon)$ have been defined together with a continuous degree density $\rho(x;\varepsilon)$.

\subsubsection{Local measures}

\begin{mydef}
The \emph{continuous $\varepsilon$-degree density}
\begin{equation}
	\rho(x;\varepsilon) = \int_{B_\varepsilon(x)} d\mu(y) \label{def:contrho}
\end{equation}
measures the probability that a point $y$ randomly drawn according to $p$ lies in an $\varepsilon$-neighborhood $B_\varepsilon(x)=\{y \in S: \|x-y\| \leq \varepsilon\}$ of $x$.
\end{mydef}

\begin{mydef}
In turn, the \emph{continuous local $\varepsilon$-clustering coefficient} of any point $x \in S$,
\begin{equation}
	\mathcal{C}(x;\varepsilon) = \frac{\int\!\!\!\int_{B_\varepsilon(x)} d\mu(y)\,d\mu(z) \Theta(\varepsilon-\|y-z\|)}{\rho(x;\varepsilon)^2}, \label{eq:local_clustering}
\end{equation}
is the probability that two points $y$ and $z$ randomly drawn according to $p$ are closer than $\varepsilon$ given they are both closer than $\varepsilon$ to $x$.
\end{mydef}

\begin{mydef}
The \emph{continuous $\varepsilon$-matching index} $\mathcal{M}(x,y,\varepsilon)$ measures the overlap of the neighborhoods of $x,y \in S$,
\begin{equation}
\mathcal{M}(x,y;\varepsilon) = \frac{\int_{B_{\varepsilon}(x) \cap B_{\varepsilon}(y)} d\mu(z)}{\int_{B_{\varepsilon}(x) \cup B_{\varepsilon}(y)} d\mu(z)}. \label{eq:matching_index}
\end{equation}
It gives the probability that a point $z$ drawn randomly from $B_{\varepsilon}(x)$ according to $p$ is also contained in $B_{\varepsilon}(y)$ and vice versa.
\end{mydef}
For $x\to y$, $\mathcal{M}(x,y;\varepsilon)\to 1$. Furthermore, $\mathcal{M}(x,y;\varepsilon)=0$ if $\|x-y\| > 2 \varepsilon$.

\subsubsection{Global measures}

\begin{mydef}
The \emph{continuous $\varepsilon$-edge density}
\begin{eqnarray}
\rho(\varepsilon) &=& \int_S d\mu(x) \rho(x;\varepsilon) \label{eq:continuous_edge_density} \\ 
&=& \int_S d\mu(x) \int_{B_\varepsilon(x)} d\mu(y) \nonumber
\end{eqnarray}
is the expectation value of the continuous $\varepsilon$-degree density $\rho(x;\varepsilon)$. 
\end{mydef}

\begin{mydef}
As a global measure of geometric transitivity, we define the \emph{continuous $\varepsilon$-transitivity} of $S$ as
\begin{eqnarray}
	\mathcal{T}(\varepsilon) &=& \bigg[ \int\!\!\!\int\!\!\!\int_S d\mu(x)\,d\mu(y)\,d\mu(z) \Theta(\varepsilon-\|x-y\|) \times \nonumber \\
&& \quad \times \Theta(\varepsilon-\|y-z\|) \Theta(\varepsilon-\|z-x\|) \bigg] \bigg/ \nonumber \\
&& \bigg[ \int\!\!\!\int\!\!\!\int_S d\mu(x)\,d\mu(y)\,d\mu(z) \Theta(\varepsilon-\|x-y\|) \times\nonumber \\ && \quad \times \Theta(\varepsilon-\|z-x\|) \bigg], \label{def:cont_transitivity}
\end{eqnarray}
which is the probability that among three points $x,y,z$ drawn randomly according to $p$, $y$ and $z$ are closer than $\varepsilon$ given they are both closer than $\varepsilon$ to $x$. 
\end{mydef}

\begin{mydef}
Similarly, the \emph{continuous global $\varepsilon$-clustering coefficient} 
\begin{equation}
\mathcal{C}(\varepsilon) = \int_S d\mu(x) \mathcal{C}(x;\varepsilon) \label{def:cont_global_clustering}
\end{equation}
is the expectation value of the continuous local $\varepsilon$-clustering coefficient $\mathcal{C}(x;\varepsilon)$~(Eq.~(\ref{eq:local_clustering})).
\end{mydef}

Note that the above defined measures of transitivity have been mainly considered for the supremum norm $L_\infty$ in~\cite{Donner2010dimensions}.

\begin{mydef}
\emph{Continuous $\varepsilon$-assortativity}
\begin{equation}
\mathcal{A}(\varepsilon) = r\bigg(\rho(x;\varepsilon),\rho(y;\varepsilon) \,\,\,\, | \,\,\,\, \|x-y\| < \varepsilon \bigg)  \label{def:cont_assortativity}
\end{equation}
gives the Pearson product-moment correlation coefficient~\cite{BrockwellDavies2002} of the degree densities $\rho(x;\varepsilon)$ and $\rho(y;\varepsilon)$ of all points $x,y$ that are closer than $\varepsilon$ to each other.
\end{mydef}
$\mathcal{A}(\varepsilon)$ can be considered as a measure of the smoothness of the set $S$ and the probability density $p$~\cite{Donner2010b}. In the limit $\varepsilon \to 0$ we have $\mathcal{A}(\varepsilon) \to 1$.

\subsection{Mesoscopic measures}

Motifs of order $\alpha$ are small connected subgraphs of $\alpha$ vertices that are embedded within the topology of a complex network~\cite{Milo2002}. For combinatorial reasons, usually $\alpha<5$ is considered. Measuring motif densities is a useful approach for quantifying higher-order neighborhood relationships in complex geometries~\cite{Xu2008} and may be seen as a generalization of the transitivity concepts introduced above.
\begin{mydef}
The \emph{continuous $\varepsilon$-motif density}
\begin{equation}
M^\alpha_\beta = \left(\prod_{i=1}^{\alpha} \int_S d\mu(x_i) \right) \prod_{(j,k) \in E^\alpha_\beta} \Theta(\varepsilon-\|x_j-x_k\|) \label{def:cont_motif_density}
\end{equation}
quantifies the frequency of occurrence of a certain recurrence motif of order $\alpha$ described by the corresponding edge set $E^\alpha_\beta$, where $\beta=1,\dots,n(\alpha)$ and $n(\alpha)$ is the total number of distinct motifs of order $\alpha$. $M^\alpha_\beta$ is the probability that $\alpha$ points drawn randomly according to $p$ are arranged according to the recurrence motif described by $E^\alpha_\beta$.
\end{mydef}
For example, the density $M^4_\sqcup$ of the recurrence motif~$\sqcup$ of order $4$ is measured by
\begin{eqnarray}
M^4_\sqcup &=& \left(\prod_{i=1}^{4} \int_S d\mu(x_i) \right) \Theta(\varepsilon-\|x_1-x_2\|) \times \nonumber \\
 && \quad \times \Theta(\varepsilon-\|x_2-x_3\|) \Theta(\varepsilon-\|x_3-x_4\|) \nonumber.
\end{eqnarray}
In contrast to the study of motifs for other general complex networks~\cite{Milo2002}, it is neither meaningful nor necessary to normalize motif densities by their expectation values for randomized networks here. The reason is that the $M^\alpha_\beta$ already have a natural probability interpretation. To render the results for different $\alpha$ more comparable, one may consider to use relative motif densities normalized by $\sum_{\beta=1}^{n(\alpha)} M_\beta^\alpha$.

We conjecture that motif densities as generalizations of the continuous $\varepsilon$-transitivity are like the latter related to certain notions of the dimensionality of the set $S$ and its associated probability density $p$~\cite{Donner2010dimensions}. This would allow us to define and study a new class of motif-based measures of dimensionality analogously to the sequence of R\'enyi dimensions from dynamical systems theory~\cite{renyi1961dimension}.  

\subsection{Path-based measures}

\begin{figure*}[tbp]
   \centering
   \includegraphics[width=\textwidth]{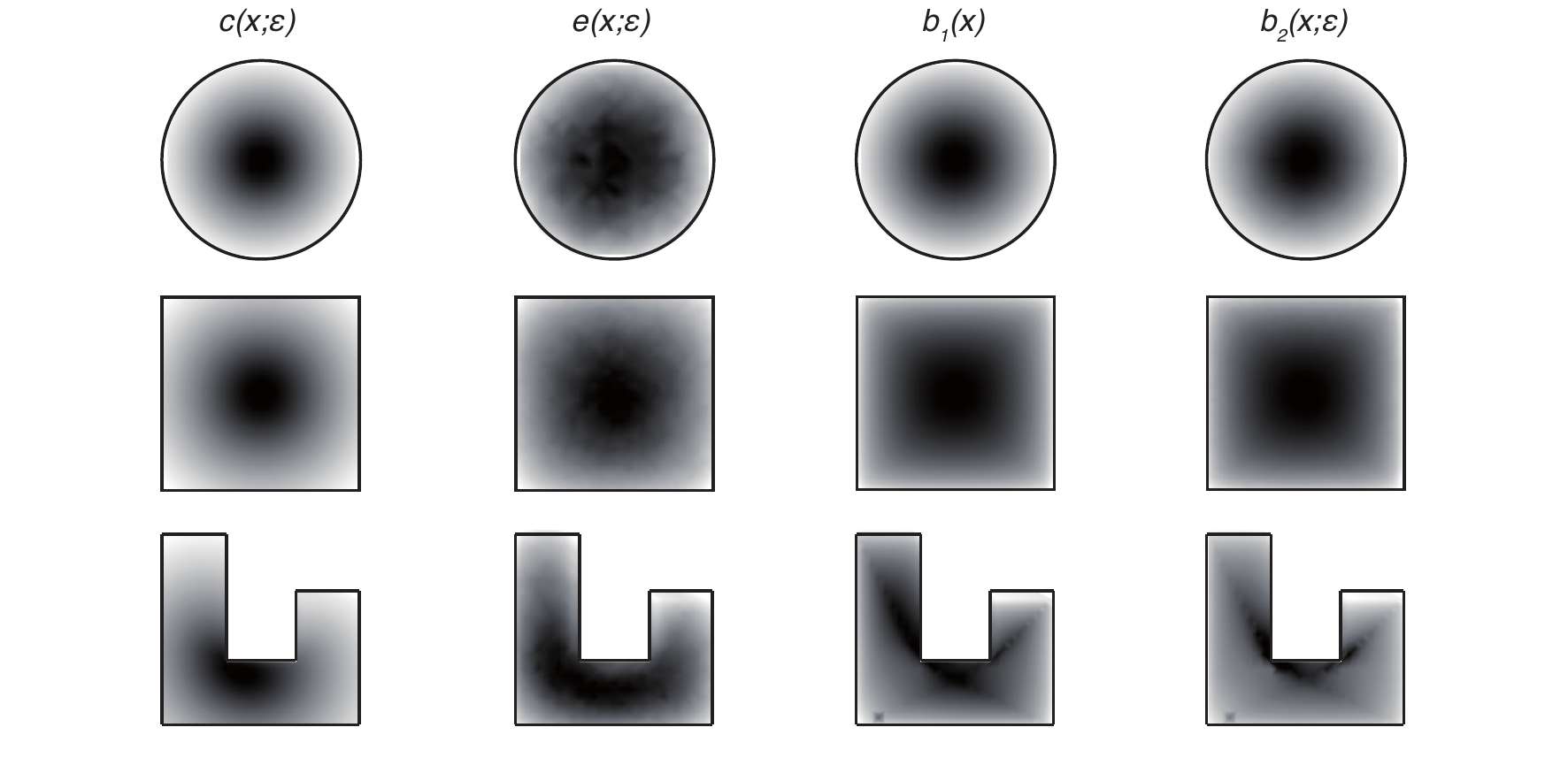}
   \caption{Local path-based measures continuous $\varepsilon$-closeness $c(x;\varepsilon)$, -efficiency $e(x;\varepsilon)$, and -shortest path betweenness $b_1(x)$ (based on Eq.~(\ref{eq:betweenness1})), $b_2(x;\varepsilon)$ (based on Eq.~(\ref{eq:betweenness2})) in three example sets $S$ with a uniform density $p$, two convex sets (circle and square), and a nonconvex set. Grayscale indicates the value of the measures (white: small, black: large) obtained by Monte Carlo numerical integration using the Euclidean norm for small $\varepsilon$ to avoid boundary effects. Note the more complex structure of the betweenness field, displaying particularly large values (dark) at the inward corners where many shortest paths must pass. In contrast to the path-based measures shown here, continuous $\varepsilon$-degree $\rho(x;\varepsilon)$ and local $\varepsilon$-clustering coefficient $\mathcal{C}(x;\varepsilon)$ are constant in the interior of $S$ due to the uniform $p$. Variations of these measures due to boundary effects occur only closer than $\varepsilon$ to the boundary of $S$~\cite{Donner2010dimensions}.}
   \label{fig:visual_examples}
\end{figure*}

While the neighborhood-based properties defined above describe the small-scale geometry of the set $S$ and probability density $p$, path-based measures quantify their global geometry in terms of global geodesics~(see Fig.~\ref{fig:visual_examples} for examples). Most of the path-based concepts defined below do not conceptually depend on the threshold $\varepsilon$. Nevertheless, we introduce the appropriate scaling with $\varepsilon$ into the definitions for consistency with the corresponding discrete estimators from complex network theory (Sec.~\ref{sec:estimators}) and the recurrence network literature.

As in standard topological terminology, a \emph{path} in $S$ is a continuous function $f : [0,1] \rightarrow S$, and its \emph{path length} $l(f) \in [0,\infty]$ is the supremum of $\sum_{i=1}^n d(f(t_{i-1}), f(t_i))$ over all $n > 0$ and all tuples $0 = t_0 \leq \dots \leq t_n = 1$, where $d(\cdot,\cdot)$ is some metric. 
Note that $l(f)$ can be infinite in which case the path is called \emph{non-rectifiable}. For points $x,y$, the \emph{geodesic distance} $g(x, y) \in [0,\infty]$ is the infimum of $l(f)$ over all paths in $S$ from $x$ to $y$ (i.e., with $f(0) = x$ and $f(1) = y$) \footnote{Note that when defined in this way, $g(x,y)$ may change discontinuously under continuous changes of the probability density $p$. This is because we require the geodesics to stay within $S$ which consists of all points $x$ where $p(x)\neq 0$. When $p(x)$ is continuously changed to zero, the length of geodesics running through $x$ for $p(x)>0$ may change abruptly once $p(x)=0$ is reached, e.g., when $x$ constitutes some kind of geometric bottleneck. If this behavior is undesirable, one may consider generalized $p$-weighted notions of the geodesic distance. These could be motivated by an analogy to the optical path length in heterogeneous and non-isotropic media in physics, where the probability density $p$ would play the role of the spatially varying refractive index.}. A corresponding path of this length is called a \emph{global geodesic} or shortest curve~\cite{oneil2006}. Depending on the geometry of $S$, there may be 
none, one, or
multiple distinct global geodesics connecting $x$ and $y$,
but in a sufficiently well-behaved set $S$, there will usually be a unique global geodesic for almost every pair $x,y$ (Fig.~\ref{fig:set_sketch}) and almost every metric (in particular, for the Euclidean metric), where by ``almost every'' we mean as usual that the set of exceptions has zero measure. Note that, however, for some pathological metrics global geodesics are rarely unique, including the $L_1$ and $L_\infty$ metrics.

To understand the reasoning behind the following definitions, one has to note that discrete shortest paths of the $\varepsilon$-recurrence network approximate global geodesics connecting two points $x,y \in S$ for small $\varepsilon$ and large $N$. Then the shortest path length $l_{ij}(\varepsilon)$ (the minimum number of edges that have to be traversed to reach vertex $i$ from vertex $j$~\cite{Newman2003}) approximates $g(x(t_i),x(t_j))$, i.e.,  
\begin{equation}
\varepsilon l_{ij}(\varepsilon) \approx g(x(t_i),x(t_j)) \label{eq:shortest_path_length}
\end{equation}
(Fig. \ref{fig:set_sketch}), where $x=x(t_i)$ and $y=x(t_j)$. In the limit $\varepsilon \to 0$, and if $N \to \infty$ sufficiently fast, we argue in Appendix~\ref{appx:proof_path_length} that indeed $\varepsilon l_{ij}(\varepsilon) \to g(x(t_i), x(t_j))$, independently of which metric is used for constructing the $\varepsilon$-recurrence network.

\subsubsection{Local measures}

\begin{mydef}
Given that a point $y$ is drawn randomly according to $p$, \emph{continuous $\varepsilon$-closeness}
\begin{eqnarray}
c(x;\varepsilon) &=& \left(\int_S d\mu(y) \frac{g(x,y)}{\varepsilon}\right)^{-1} \label{def:cont_closeness} \\
&=& \varepsilon \left(\int_S d\mu(y) g(x,y)\right)^{-1} \nonumber
\end{eqnarray}
is the inverse expected geodesic distance of $y$ to another chosen point $x$ in units of $\varepsilon$. 
\end{mydef}

\begin{mydef}
Similarly, \emph{continuous local $\varepsilon$-efficiency} 
\begin{eqnarray}
e(x;\varepsilon) &=& \int_S d\mu(y) \left(\frac{g(x,y)}{\varepsilon}\right)^{-1} \label{def:cont_efficiency} \\
&=& \varepsilon \int_S d\mu(y) g(x,y)^{-1} \nonumber
\end{eqnarray}
gives the expected inverse geodesic distance of $y$ to $x$ measured in units of $\varepsilon$.
\end{mydef}
Both $c(x;\varepsilon)$ and $e(x;\varepsilon)$ quantify the geometric closeness of $x$ to any other point in $S$ given a probability density $p$. Hence, points in the center of $S$ will carry larger values of $c(x;\varepsilon)$ and $e(x;\varepsilon)$ than those on its boundaries (see Fig.~\ref{fig:visual_examples} and below).

\begin{figure}[tbp]
   \centering
   \includegraphics[width=\columnwidth]{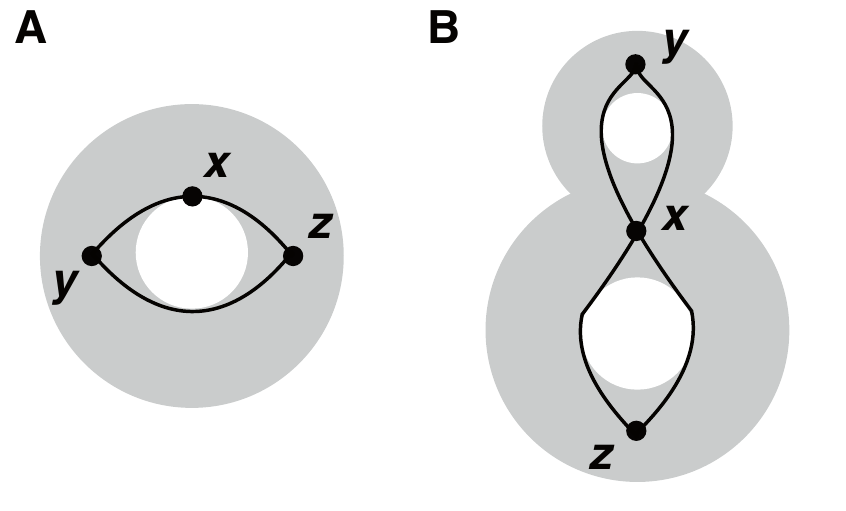}
   \caption{Illustration of the definition of continuous $\varepsilon$-shortest path betweenness (the set $S$ is indicated by gray shading). (A) There are $\sigma(y,z)=2$ global geodesics connecting $y,z \in S$, but only $\sigma(y,z;x;\varepsilon)=1$ includes $x \in S$. (B) In this example, $x$ lies on all four global geodesics between $x$ and $y$, i.e., $\sigma(y,z)=\sigma(y,z;x;\varepsilon)=4$.}
   \label{fig:betweenness_sketch}
\end{figure}

\begin{mydef}
\emph{Continuous $\varepsilon$-shortest path betweenness}
\begin{equation}
b(x;\varepsilon) = \int\!\!\!\int_S d\mu(y)\,d\mu(z) \frac{\sigma(y,z;x;\varepsilon)}{\sigma(y,z)}, \label{def:cont_sp_betweenness}
\end{equation}
is the probability that a point $x$ lies on a randomly chosen global geodesic connecting two points $y,z$ drawn randomly from $S$ according to $p$. Here, $\sigma(y,z;x;\varepsilon)$ denotes the number of times $x \in S$ lies on a global geodesic between $y,z \in S$ and $\sigma(y,z)$ is the total number of global geodesics between $y,z$ (Fig.~\ref{fig:betweenness_sketch}).
\end{mydef}

In pathological situations, e.g., for certain open sets $S$, $\sigma(y,z)$ may be zero even when the geodesic distance $g(y,z)$ is well-defined and finite. We ignore these cases for now.

There are several ways to formally define $\sigma(y,z;x;\varepsilon)$. Using a parametrization $f_\kappa(t)$ of the family of global geodesics connecting $y$ and $z$, with $t \in [0,1]$ and $f_\kappa(0) = y$, $f_\kappa(1) = z$, we may write
\begin{eqnarray}
\sigma_1(y,z;x;\varepsilon) &=& \sigma_1(y,z;x) \nonumber \\
&=& \sum_{\kappa=1}^{\sigma(y,z)} \int_0^1 dt \: \delta(f_\kappa(t) - x), \label{eq:betweenness1}
\end{eqnarray}
where $\delta(\cdot)$ is Dirac's multi-dimensional delta function. Alternatively, we can include the finite $\varepsilon$-effect by counting all shortest paths that pass through the $\varepsilon$-neighborhood of $x$ by setting
\begin{equation}
\sigma_2(y,z;x;\varepsilon) = \sum_{\kappa=1}^{\sigma(y,z)} \int_0^1 dt \: \Theta(\varepsilon - \|f_\kappa(t) - x\|). \label{eq:betweenness2}
\end{equation}
Both variants of $\sigma(y,z;x;\varepsilon)$ yield different, yet qualitatively similar results for $b(x;\varepsilon)$ as is illustrated in Fig.~(\ref{fig:visual_examples}).
 
Given convex domains $S$, $\sigma(y,z)=1$ always holds, i.e., there is only one straight line connecting $y$ and $z$, parametrized by $f(t)=y + t (z-y)$. For one-dimensional, convex sets $S$ and using $\sigma_1(y,z;x)$, continuous $\varepsilon$-shortest path betweenness simplifies to
\begin{equation}
b(x) = 2 \int\!\!\!\int_S d\mu(y)\,d\mu(z) \Theta(x-y) \Theta(z-x).
\end{equation}

\begin{mydef}
\emph{Continuous $\varepsilon$-shortest path edge betweenness}
\begin{equation}
b(x,y;\varepsilon) = \int\!\!\!\int_S d\mu(z)\,d\mu(w) \frac{\sigma(z,w;x,y;\varepsilon)}{\sigma(z,w)}, \label{eq:edge_betweenness}
\end{equation}
is the probability that two points $x,y$ both lie on a randomly chosen global geodesic connecting two points $z,w$ drawn randomly according to $p$. $\sigma(z,w;x,y;\varepsilon)$ counts the number of global geodesics between $z,w$ which contain $x,y$. 
\end{mydef}
Analogously to continuous $\varepsilon$-shortest path betweenness $b(x;\varepsilon)$, we can define this quantity as
\begin{eqnarray}
\sigma_1(z,w;x,y;\varepsilon) &=& \sigma_1(z,w;x,y) \nonumber \\
&=& \sum_{\kappa=1}^{\sigma(z,w)} \left( \int_0^1 dt \: \delta(f_\kappa(t) - x)\right) \times \nonumber \\
&& \qquad \times \left( \int_0^1 dt \: \delta(f_\kappa(t) - y) \right).
\end{eqnarray}
Further generalizations for including the finite $\varepsilon$-effect may be deduced as shown above for continuous $\varepsilon$-shortest path betweenness. 

For one-dimensional convex sets $S$ and using $\sigma_1(z,w;x,y;\varepsilon)$, Eq.~(\ref{eq:edge_betweenness}) reduces to
\begin{eqnarray}
b(x,y) &=& 2 \int\!\!\!\int_S d\mu(z)\,d\mu(w) \Theta(x-z) \Theta(y-z) \times \nonumber \\
&& \qquad \times \Theta(w-x) \Theta(w-y).
\end{eqnarray}
In the limit $x \to y$ we always have $b(x,y;\varepsilon) \to b(x;\varepsilon)$. We note that $b(x,y;\varepsilon)$ does not require the condition $\Theta(\varepsilon - \|x-y\|)$ as is the case for the corresponding discrete estimator (Table \ref{tab:estimators}). Related generalized concepts of co- and group betweenness have been described for discrete complex networks \cite{kolaczyk2009group}.

For general non-pathological $S$ we almost surely have $\sigma(z,w)=1$, i.e., the probability that there are more than one global geodesics connecting $z$ and $w$ drawn randomly from $S$ according to $p$ is zero. For example, in both Figs.~\ref{fig:betweenness_sketch}A and B the set of pairs $z,w$ with $\sigma(z,w)=2$ or $\sigma(z,w)=4$, respectively, is of measure zero. In these cases, $b(x;\varepsilon)$ and $b(x,y;\varepsilon)$ reduce to
\begin{equation}
b(x;\varepsilon) = \int\!\!\!\int_S d\mu(y)\,d\mu(z) \sigma(y,z;x;\varepsilon)
\end{equation}
and
\begin{equation}
b(x,y;\varepsilon) = \int\!\!\!\int_S d\mu(z)\,d\mu(w) \sigma(z,w;x,y;\varepsilon).
\end{equation}

It should be noted that for general $S$ and $p$, the center of mass
\begin{equation}
X = \int_S d\mu(x) x
\end{equation}
does not necessarily extremize $c(x;\varepsilon)$, $e(x;\varepsilon)$ or $b(x)$. However, for convex $S$ the generalized continuous $\varepsilon$-closeness
\begin{eqnarray}
c_\eta(x;\varepsilon) &=& \left(\int_S d\mu(y) \left(\frac{g(x,y)}{\varepsilon}\right)^\eta \right)^{-1} \\
&=& \varepsilon^\eta \left(\int_S d\mu(y) g(x,y)^\eta \right)^{-1} \nonumber
\end{eqnarray}
can be shown to assume a global maximum at $x=X$ for the special case $\eta=2$. In turn, the standard continuous $\varepsilon$-closeness $c(x;\varepsilon)$ ($\eta=1$) is maximized at the geometric median or Fermat-Weber point~\cite{Fekete2005}.

\subsubsection{Global measures}

\begin{mydef}
The \emph{continuous $\varepsilon$-average path length}
\begin{eqnarray}
\mathcal{L}(\varepsilon) &=& \int\!\!\!\int_S d\mu(x)\,d\mu(y) \frac{g(x,y)}{\varepsilon} \label{eq:apl} \\
&=& \varepsilon^{-1} \int\!\!\!\int_S d\mu(x)\,d\mu(y) g(x,y). \nonumber
\end{eqnarray}
measures the expected geodesic distance in units of $\varepsilon$ between two points $x$ and $y$ drawn randomly according to $p$. 
\end{mydef}
From Eq.~(\ref{eq:apl}), the equivalence of this formulation of continuous average path length to the intensively studied problem in probabilistic geometry~\cite{Bailey2006} of finding the expectation value of the distance between two randomly drawn points $x,y\in S$ according to the probability distribution $p$ becomes evident.
Our definitions imply the relationship
\begin{equation}
\mathcal{L}(\varepsilon) = \int_S d\mu(x) c(x;\varepsilon)^{-1}.
\end{equation}

\begin{mydef}
Similarly, the \emph{continuous global $\varepsilon$-efficiency}
\begin{eqnarray}
\mathcal{E}(\varepsilon) &=& \left(\int\!\!\!\int_S d\mu(x)\,d\mu(y) \left(\frac{g(x,y)}{\varepsilon}\right)^{-1}\right)^{-1} \label{eq:global_efficiency} \\
&=& \varepsilon^{-1} \left(\int\!\!\!\int_S d\mu(x)\,d\mu(y) g(x,y)^{-1}\right)^{-1}. \nonumber
\end{eqnarray}
is the inverse of the expected inverse geodesic distance between two points $x,y$ drawn randomly according to $p$ measured in units of $\varepsilon$. 
\end{mydef}
Here, we have
\begin{equation}
\mathcal{E}(\varepsilon) = \left(\int_S d\mu(x) e(x;\varepsilon)\right)^{-1}.
\end{equation}
More generally, let $\Delta_{S,p}(\eta)$ be the expectation value of a power $\eta$ of the geodesic distance $g(x,y)$ between two points $x,y\in S$ randomly drawn according to $p$:
\begin{equation}
\Delta_{S,p}(\eta) = \int\!\!\!\int_S d\mu(x)\,d\mu(y) g(x,y)^{\eta}. \label{eq:delta_def}
\end{equation}
Then continuous $\varepsilon$-average path length and global $\varepsilon$-efficiency may be expressed as
\begin{equation}
\mathcal{L}(\varepsilon) = \varepsilon^{-1} \Delta_{S,p}(1)
\end{equation}
and
\begin{equation}
\mathcal{E}(\varepsilon) = \varepsilon^{-1} \left(\Delta_{S,p}(-1)\right)^{-1}.
\end{equation}

\begin{mydef}
The \emph{$\varepsilon$-diameter}
\begin{equation}
\mathcal{D}(\varepsilon) =  \varepsilon^{-1}  \sup_{x,y \in S} g(x,y) \label{eq:diameter}
\end{equation}
and the \emph{$\varepsilon$-radius} 
\begin{equation}
\mathcal{R}(\varepsilon) =  \varepsilon^{-1}  \inf_{x \in S} \sup_{y \in S} g(x,y) \label{eq:radius}
\end{equation}
are global geometric characteristics of the set $S$ that are independent of $p$ \cite{oneil2006}.
\end{mydef}

\subsection{Further measures}

To illustrate that the proposed framework can be extended in several directions, we shortly discuss spectral and random-walk-based measures in the context of continuous recurrence networks. Motivated by the study of eigenvector centrality in complex networks~\cite{Newman2003}, we can consider spectral properties of the set $S$ and probability density $p$.
\begin{mydef}
The linear Laplace operator
\begin{eqnarray}
\left(\mathbb{L}_\varepsilon f\right)(x) &=& \int_S d\mu(y) \large[\Theta(\varepsilon-\|x-y\| \nonumber \\
&& \qquad\qquad - \delta(x-y) \rho(y;\varepsilon) \large] f(y)
\end{eqnarray}
is a continuous equivalent of the discrete Laplacian matrix in network theory~\cite{Newman2003}. We are interested in its eigenfunctions $f(x)$ and eigenvalues $\lambda$ satisfying
\begin{equation}
\left(\mathbb{L}_\varepsilon f\right)(x) = \lambda f(x)
\end{equation}
for all $x\in S$.
\end{mydef}
For example, considering an arbitrary $S$ with uniform~$p$, one obtains an eigenfunction $f(x)=C$ for some $C\in \mathbb{R}$ associated to the eigenvalue $\lambda=0$. This is analogous to the eigenvector $(1,1,\dots,1)$ with eigenvalue $0$ which is always present for the discrete Laplacian matrix of general networks~\cite{Newman2003}. We can expect more interesting results for non-uniform $p$. For example, one may define a continuous analog of the eigenvector centrality of complex network theory~\cite{Bonacich1972} by considering the eigenfunction $\tilde{f}(x)$ corresponding to the largest eigenvalue $\tilde{\lambda}$.

For discrete networks, there are several measures of betweenness based on random walks rather than shortest paths \cite{Arenas2003,Newman2005}.
Continuous versions of these measures would be based on continuous analogs of random walks on $S$ that start and end at points $y$ and $z$ randomly chosen from $p$. 
Since in a discrete network the limit distribution of a random walk without a sink is proportional to the degree distribution, a natural choice for a continuous analog is an It\= o diffusion process, the limit distribution of which is proportional to $\rho(x;\varepsilon)$, with a source at $y$ and a sink at $z$~\cite{Oksendal2003}.
Such a process can most easily be defined as a gradient flow 
$dX_t = -\nabla\Psi(X_t)\,dt + \sqrt{2T}\,dB_t$
that combines a Brownian motion $B$ with a local drift coefficient $-\nabla\Psi(X_t)$ which comes from a potential $\Psi(x)$ that is the product of a temperature $T>0$ and the information corresponding to $\rho(x;\varepsilon)$, which is $-\ln\rho(x;\varepsilon)$. The resulting process
\begin{eqnarray}
dX_t = T\frac{\nabla \rho(X_t;\varepsilon)}{\rho(X_t;\varepsilon)}\,dt + \sqrt{2T}\,dB_t
\end{eqnarray}
can then be interpreted as a diffusion that drifts in the direction of increasing density.
The continuous version of Arenas' random walk betweenness \cite{Arenas2003} would then be the expected density of the process at $x$ when the source and sink are drawn from $p$.
Similarly, the continuous version of Newman's random walk betweenness \cite{Newman2005} would be the expected absolute value of the resulting flux density at $x$ for a random source and sink.

\subsection{Behavior under affine transformations}
\label{sec:affine_trafos}

All continuous measures defined above are based on neighborhood relationships in $S$ and geodesic distances between points therein. They are therefore invariant with respect to the subclass of affine transformations which leaves these properties unchanged, i.e., $x \rightarrow Dx+s$ for $x\in S$ with $D$ being a combination of rotation and isotropic scaling operations and $s$ a translation. This is to be understood in the sense that for a measure $h$, $h(Dx+s; a \varepsilon)=h(x; \varepsilon)$ holds, where $a$ is the scaling factor. The measures considered here are generally not invariant under non-isotropic scaling and shear operations.

\section{Discrete estimators}
\label{sec:estimators}

\begin{table*}[htdp]
\caption{A summary of standard unweighted network estimators for the continuous geometric properties defined in Sec. \ref{sec:continuous_framework}. For a detailed discussion, see~\cite{Donner2010b,Newman2003,Boccaletti2006,costa2007}. SP abbreviates ``shortest path".}
\label{tab:estimators}
\begin{center}
\begin{tabular}{llll}
Class & Name & Definition & Comments\\
\hline
\textit{Neighborhood-based} & & & \\
Local & Degree & $\hat{k}_i = \sum_{j=1}^{N} A_{ij}$ & \\
& Degree density & $\hat{\rho}_i = \frac{1}{N-1} \hat{k}_i$ & \\
& Local clustering coeff. & $\hat{\mathcal{C}}_i = \frac{\sum_{j,k=1}^N A_{ij}A_{jk}A_{ki}}{\hat{k}_i (\hat{k}_i-1)}$ & $\hat{\mathcal{C}}_i = 0$ iff $k_i < 2$. \\
& Matching index & $\hat{\mu}_{ij} = \frac{\sum_{l=1}^N A_{il} A_{jl}}{\hat{k}_i + \hat{k}_j - \sum_{l=1}^N A_{il} A_{jl}}$ & \\
& & & \\
Global & Edge density & $\hat{\rho} = \frac{1}{N(N-1)}\sum_{i,j=1}^N A_{ij}$ & \\
& Transitivity & $\hat{\mathcal{T}} = \frac{\sum_{i,j,k=1}^N A_{ij}A_{jk}A_{ki}}{\sum_{i,j,k=1}^N A_{ki}A_{kj}}$ & \\
& Global clustering coeff. & $\hat{\mathcal{C}} = \frac{1}{N} \sum_i \hat{\mathcal{C}}_i$ & \\
& Assortativity & $\hat{\mathcal{A}} = \frac{\frac{1}{L} \sum_{j>i} \hat{k}_i \hat{k}_j A_{ij} - \left<\frac{1}{2} (\hat{k}_i+\hat{k}_j)\right>_{i,j}^2}{\frac{1}{L} \sum_{j>i} \frac{1}{2} (\hat{k}_i^2+\hat{k}_j^2)A_{ij} - \left<\frac{1}{2} (\hat{k}_i+\hat{k}_j)\right>_{i,j}^2}
$ & $L=\sum_{j>i} A_{ij}$ is the number of edges, \\
& & & $\left<\frac{1}{2} (\hat{k}_i+\hat{k}_j)\right>_{i,j} =\frac{1}{L} \sum_{j>i} \frac{1}{2} (\hat{k}_i+\hat{k}_j)A_{ij}$~\cite{Donges2011b}. \\
\hline
\textit{Path-based} & & & \\
Local & Closeness & $\hat{c}_i = \frac{N-1}{\sum_{j=1}^N l_{ij}}$ & Set $l_{ij}=N-1$ iff $\nexists$ path between $i,j$~\cite{Freeman1977}. \\
& Local efficiency & $\hat{e}_i = \frac{1}{N-1}{\sum_{j=1}^N l_{ij}^{-1}}$ & \\
& SP betweenness & $\hat{b}_i = {N-1 \choose 2}^{-1} \sum_{j,k\neq i}^N \frac{\hat{\sigma}_{jk}(i)}{\hat{\sigma}_{jk}}$ & $\hat{\sigma}_{jk}$ shortest paths connect vertices $j,k$, \\
& & & $\hat{\sigma}_{jk}(i)$ of those include $i$~\cite{Freeman1977}, \\
& SP edge betweenness & $\hat{b}_{ij} = {N-1 \choose 2}^{-1} \sum_{k,l\neq i,j}^N \frac{\hat{\sigma}_{kl}(i,j)}{\hat{\sigma}_{kl}}$ & and $\hat{\sigma}_{jk}(i,j)$ include $i,j$. \\
Global & Average path length & $\hat{\mathcal{L}} = \left<l_{ij}\right>_{i,j}$ & Set $l_{ij}=N-1$ iff $\nexists$ path between $i,j$~\cite{Newman2003}. \\
& Global efficiency & $\hat{\mathcal{E}} = \left(\left<l_{ij}^{-1}\right>_{i,j}\right)^{-1}$ & \\
& Diameter & $\hat{\mathcal{D}} = \max_{i,j} \left(l_{ij}\right)$ & \\
& Radius & $\hat{\mathcal{R}} = \min_i \max_j \left(l_{ij}\right)$ & \\
\end{tabular}
\end{center}
\label{default}
\end{table*}

Given the continuous framework defined above, we are able to treat the commonly used recurrence network quantifiers~\cite{Donner2010b,Donner2011a} taken from standard complex network theory~\cite{Newman2003,Boccaletti2006} as the most straightforward \emph{discrete estimators} of the continuous quantities for a finite number of observations $N$ and finite $\varepsilon$. The discrete estimators will be denoted using hats, e.g., the discrete estimator of continuous average path length $\mathcal{L}(\varepsilon)$ is $\hat{\mathcal{L}}(\varepsilon,N)$ (we will in the following omit the estimators' dependency on $\varepsilon$ and $N$ to simplify the notation). Their numerical properties have been elaborated in detail in earlier works~\cite{Donner2010a,Donner2010b,Donner2010dimensions,Donner2011a,Marwan2009,Zou2010}. The characteristics of these standard measures for discrete and finite complex networks have also been studied for random geometric graphs and more general network models with strong spatial contraints~\cite{Barnett2007,Herrmann2003,Itzkovitz2005}, e.g., the degree distribution~\cite{Herrmann2003}, network motifs~\cite{Itzkovitz2005}, as well as clustering coefficient and degree correlations~\cite{Barnett2007}.

Here, we briefly review the estimator's definitions (Table \ref{tab:estimators}). For some specific examples, the estimators will be compared to the results theoretically derived from their continuous counterparts in Sec.~\ref{sec:examples}. This will also allow us to gain certain insights into their bias and variance for finite data sets.

\subsection{Weighted network statistics and node splitting invariant measures}

We may now ask how the estimation of the above defined continuous geometric properties 
from a finite data set can be improved with respect to the measures from complex network theory that have been used so far for this purpose. 
One way to go in line with standard estimation theory is node-weighted network statistics, 
as proposed by Heitzig \textit{et al.}~\cite{Heitzig2011}. 
For a full application of that theory, weights $w_i$ for all vertices $i$ have to be chosen in a suitable way, 
which we leave as a subject of future research.
But even with constant weights $w_i\equiv 1$, 
the axiomatic theory developed in \cite{Heitzig2011} allows us to improve estimation
by using so-called {\em node splitting invariant (n.s.i.)} versions of network measures
to reduce the estimation bias that results from excluding self-loops from the network.
Let us illustrate this for the case of continuous $\varepsilon$-degree density,
$\rho(x;\varepsilon) = \int_{B_\varepsilon(x)} d\mu(y)$.
If $x$ is a vertex, $p$ is approximately constant in $B_\varepsilon(x)$,
and the latter contains $\hat{k}_i$ additional vertices (see Table \ref{tab:estimators}), then 
$\rho(x;\varepsilon) \approx p(x)\text{Vol}(B_\varepsilon(x)) \approx (\hat{k}_i + 1)/N$
since $B_\varepsilon(x)$ contains $\hat k^\ast_i=\hat{k}_i + 1$ out of $N$ vertices.
In other words, $\hat k^\ast_i/N$ is a better estimator for $\rho(x;\varepsilon)$ than
$\hat k_i/N$, since the latter has a bias of $\sim\mathcal{O}(1/N)$.
Likewise, the transitivity measure $\hat{\mathcal{T}}$ (see Table \ref{tab:estimators}) 
can be improved by using instead the n.s.i.\ transitivity
$\hat{\mathcal{T}}^\ast = \sum_{i,j,k=1}^N A^+_{ij}A^+_{jk}A^+_{ki}/\sum_{i,j,k=1}^N A^+_{ki}A^+_{kj}$,
where $A^+_{ij} = A_{ij} + \delta_{ij} = R_{ij}$, 
showing that this approach is also more in line with recurrence plot analysis.
This would also reduce the bias in the estimation of the {\em transitivity dimension}
that was observed in \cite[Fig.\,10A]{Donner2010dimensions}.
The measures $\hat k^\ast_i$ and $\hat{\mathcal{T}}^\ast$ are examples of n.s.i.\ measures with unit weights, 
which can basically be interpreted as 
variants of the classical measures in which vertices are considered to be linked to themselves \cite{Heitzig2011}.

\subsection{Choice of the recurrence threshold $\varepsilon$}
\label{sec:recurrence_threshold}

A careful choice of the recurrence threshold $\varepsilon$ is critical for faithfully estimating the continuous recurrence network properties defined above~\cite{Donner2010a}. For too large $\varepsilon$, i.e., on the order of the diameter of $S$, boundary effects dominate, the discrete recurrence network used for estimation becomes too dense and is unable to capture the geometry induced by $S$ and $p$ (see Sec.~\ref{sec:comparison_numerical}). In contrast, the network's giant component breaks down for too small $\varepsilon$ with a phase transition at the critical value $\varepsilon_c$. This obstructs our ability to properly estimate mesoscopic and path-based measures for $\varepsilon<\varepsilon_c$. Therefore we expect a good performance of the discrete estimators for thresholds just above the critical $\varepsilon_c$, where much of the geometric fine structure is still resolved~\cite{Donner2010dimensions}.

The problem of selecting $\varepsilon$ therefore reduces to deriving the percolation threshold $\varepsilon_c$ which is directly related to the critical edge density $\rho_c=\rho(\varepsilon_c)$ of the theory of random geometric graphs~\cite{Penrose2003} via Eq.~(\ref{eq:continuous_edge_density}). $\rho_c$ is linked to the commonly studied critical mean degree $z_c$ by
\begin{equation}
\rho_c = \frac{z_c}{N-1}.
\end{equation} 
The Erd\H{o}s-R\'enyi graph is the simplest random network model~\cite{Newman2003}. Since any pair of vertices is linked with the same probability $\rho$ independent of their distance, it neglects the effects of spatial embedding. Therefore the Erd\H{o}s-R\'enyi model is inadequate for describing $d$-dimensional random geometric graphs, and the corresponding critical mean degree $z_c=1$~\cite{Newman2003} turns out to be too low except for the limiting case $d\to \infty$~\cite{Dall2002} (see Sec.~\ref{sec:comparison_numerical} for an example). Taking into account the effects of clustering of vertices induced by the spatial embedding \cite{Bialonski2010} yields improved analytical bounds on the true $z_c$ obtained from numerical simulations~\cite{Kong2007}. Exact analytical results for arbitrary $d$ are not available so far, but Dall and Christensen~\cite{Dall2002} have empirically found the scaling law
\begin{equation}
z_c(d) = z_c(\infty) + A d^{-\gamma}
\end{equation}
from extensive numerical simulations, where $z_c(\infty)=1$, $\gamma=1.74(2)$ and $A=11.78(5)$. 
Inverting $\rho_c=\rho(\varepsilon_c)$ (which is possible as $d\rho(\varepsilon)/d\varepsilon>0$ in non-pathological situations) yields the associated critical threshold
\begin{equation}
\varepsilon_c(d) = \rho^{-1}\left(\frac{z_c(d)}{N-1}\right). \label{eq:critical_threshold}
\end{equation}
To our best knowledge this is the most useful result available so far for our aim of choosing the recurrence threshold $\varepsilon$. However, one should be aware that the results of~\cite{Dall2002} were obtained for the box $S=[0,1]^d$ with uniform probability density $p$ which is the most commonly studied setting in random geometric graph theory. When considering general $S$ and $p$ they may be appropriate as a first educated guess for properly selecting $\varepsilon$ in line with the guidelines discussed in~\cite{Donner2010a,Donner2010dimensions}. Deriving analytical bounds on $z_c$ for such geometries remains an open problem.

\section{Examples} 
\label{sec:examples}

We illustrate the above defined continuous geometric quantities and their estimators for paradigmatic examples by giving closed-form analytical results and relating them to numerical evidence from $\varepsilon$-recurrence networks constructed from time series. The focus will be on examples where all quantities of interest can be calculated either analytically or semi-analytically (relying on numerical evaluation of some integrals), i.e., possessing smooth sets $S$ and density functions $p(x)$. This implies that when considering the Euclidean norm (which we will use for all examples below) and neglecting boundary effects~\cite{Donner2010b,Donner2010dimensions}, we obtain
\begin{equation}
\begin{split}
\mathcal{C}(x;\varepsilon) = & \ 1 - \frac{d\Gamma(d/2)}{2\sqrt\pi\Gamma((d+1)/2)}
\bigg[{}_2F_1\left(\frac 1 2,\frac{1-d}2;\frac 3 2; \frac 1 4\right) \\
& \qquad - \frac 1{d+1} {}_2F_1\left(\frac{1-d}2,\frac{d+1}2;
				\frac{d+3}2;\frac 1 4\right) \bigg]\\
= & \mathcal{C}(\varepsilon) = \mathcal{T}(\varepsilon) \label{eq:hypergeom_transitivity}
\end{split}
\end{equation}
for all transitivity-based measures, where ${}_2F_1(\cdot)$ is the hypergeometric function and $d$ the manifold dimension of $S$. A simpler exponential scaling with $d$ can be found for the supremum metric. Nontrivial transitivity-based properties for fractal sets $S$ and densities $p(x)$ allowing for non-integer $d$, where an analytical calculation of path-based measures is problematic, have been treated exhaustively in~\cite{Donner2010dimensions}.

The results given here hold in the limit $\varepsilon \to 0$. For simplicity we ignore boundary effects which have been treated in~\cite{Donner2010b,Donner2010dimensions}. In all examples, we use the parametrization $\sigma_1(y,z;x)$ to compute continuous $\varepsilon$-shortest path betweenness.

\subsection{One-dimensional chaotic maps and stochastic processes}

All examples considered in this section are defined on convex sets $S$ embedded on the real axis. Therefore, the geodesic distance of $x,y\in \mathbb{R}$ reduces to $g(x,y)=|x-y|$. Since for one-dimensional $S$ the integral $\int_S\,dy\,p(y)\,|x-y|^{-1}$ diverges for all $p$ and all $x\in S$, we get $\mathcal{E}(\varepsilon) = 0$ and $e(x;\varepsilon)=\infty$ $\forall x$ in all examples of this section. In contrast, the corresponding integral always converges for non-fractal $S$ with $d \geq 2$ and general $p$.

\subsubsection{Bernoulli map / uniformly distributed noise} 

The Bernoulli map $x_{n+1}=\left(2 x_n\right)\,\mod\,1$ defined on the interval $S=[0,1)$ induces the probability density $p(x)=1$. This yields
\begin{eqnarray*}
\mathcal{T}(\varepsilon)&=&\mathcal{C}(\varepsilon) = \mathcal{C}(x;\varepsilon) = \frac{3}{4} \\
\mathcal{L}(\varepsilon) &=& \frac{1}{3} \varepsilon^{-1} \\
c(x;\varepsilon) &=& \frac{2 \varepsilon}{1-2x+2x^2} \\
b(x) &=& 2x(1-x).
\end{eqnarray*}
The same results hold for uniformly distributed noise on the interval $[0,1]$, since $S$ and $p(x)$ are identical to those of the Bernoulli map (an exemplary calculation for this setting is shown in Appendix~\ref{appx:bernoulli_map}). This equality clearly illustrates that recurrence network analysis is purely geometric and, hence, by design masks out the auto-dependency structure of dynamical systems. Stochastic and deterministic dynamics can be distinguished when embedding techniques are used prior to recurrence network analysis~\cite{Donner2010b}.

\subsubsection{Gaussian noise}

Considering Gaussian noise with zero mean and standard deviation $\sigma$, i.e., $p(x)=\left(1/\sqrt{2 \pi \sigma^2}\right) \exp{\left(-x^2 / (2 \sigma^2)\right)}$, on the real axis $S=(-\infty,+\infty)$ we obtain
\begin{eqnarray*}
\mathcal{T}(\varepsilon)&=&\mathcal{C}(\varepsilon) = \mathcal{C}(x;\varepsilon) = \frac{3}{4} \\
\mathcal{L}(\varepsilon) &=& \frac{2 \sigma}{\sqrt{\pi}} \varepsilon^{-1} \\
c(x;\varepsilon) &=& \frac{\varepsilon }{\sqrt{\frac{2}{\pi }} \sigma \exp\left(-\frac{x^2}{2 \sigma ^2}\right) + x \text{erf}\left(\frac{x}{\sqrt{2} \sigma }\right)} \\
b(x) &=& \frac{1}{2} \left(1-\text{erf}\left(\frac{x}{\sqrt{2} \sigma }\right)^2\right),
\end{eqnarray*}
where $\text{erf}(x)=\frac{2}{\sqrt{\pi }}\int _0^xe^{-t^2}dt$ is the error function. The results for mean $\chi \neq 0$ can be derived by substituting $x \rightarrow x - \chi$ on the right side of the equations for the local measures given above (see also Sec.~\ref{sec:affine_trafos}).

\subsubsection{Logistic map}

We can also give exact analytical solutions for the logistic map in the fully chaotic regime, $x_{n+1}=4x_n(1-x_n)$, defined on the interval $S=[0,1]$. Using the probability density $p(x)=\pi^{-1}\sqrt{x(1-x)}^{-1}$~\cite{Yorke_book} yields
\begin{eqnarray*}
\mathcal{T}(\varepsilon)&=&\mathcal{C}(\varepsilon) = \mathcal{C}(x;\varepsilon) = \frac{3}{4} \\
\mathcal{L}(\varepsilon) &=& \frac{4}{\pi^2} \varepsilon^{-1} \\
c(x;\varepsilon) &=& \pi \varepsilon \big[ 2\sqrt{x(1-x)} + (1-2x) \times \\ 
& & \qquad \times \left(\arccos\left(\sqrt{x}\right) - \arcsin\left(\sqrt{x}\right)\right) \big]^{-1} \\
b(x) &=& \frac{8 \: \text{Im}\left[\text{arcosh}\left(\sqrt{x}\right)\right] \arcsin\left(\sqrt{x}\right)}{\pi ^2} \nonumber.
\end{eqnarray*}

\subsubsection{Comparison to numerical results}
\label{sec:comparison_numerical}

\begin{figure}[tbp]
   \centering
   \includegraphics[width=0.49\columnwidth]{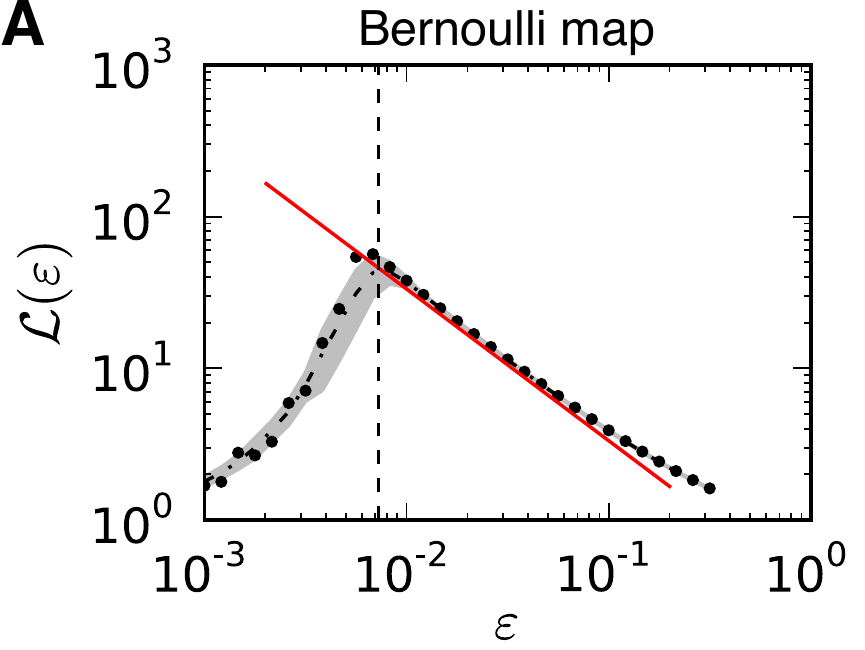}
   \includegraphics[width=0.49\columnwidth]{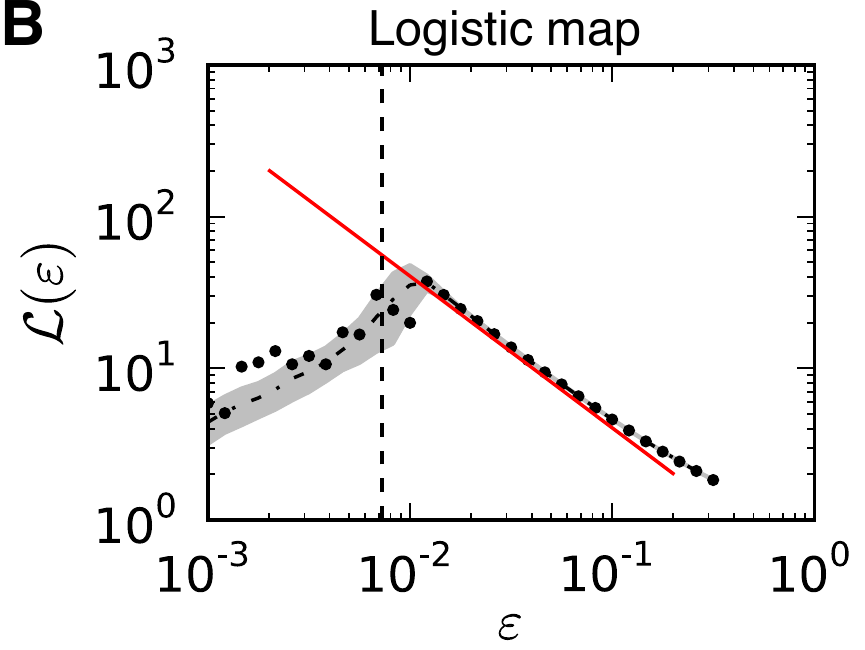}
   \caption{(Color online) Continuous $\varepsilon$-average path length $\mathcal{L}(\varepsilon)$ for (A) the Bernoulli map and (B) the logistic map. Analytical results are indicated by solid red lines. Estimates $\hat{\mathcal{L}}(\varepsilon)$ have been obtained from $\varepsilon$-recurrence networks constructed from one realization of $N=1,000$ samples, respectively, for each map (black dots). Ensemble mean (dashed-dotted black line) and standard deviation (gray band) for different $\varepsilon$ have been obtained from an ensemble of 100 realizations of each model with initial conditions uniformly distributed in the interval $[0,1]$. Vertical dashed lines indicate the estimated percolation thresholds $\varepsilon_c$.}
\label{fig:one_dim_map_apl}
\end{figure}

Within an intermediate range of $\varepsilon$, the continuous $\varepsilon$-average path length $\mathcal{L}(\varepsilon)$ is approximated well by the estimators $\hat{\mathcal{L}}(\varepsilon)$ calculated from $\varepsilon$-recurrence networks for both the Bernoulli and logistic map (Fig.~\ref{fig:one_dim_map_apl}). For small $\varepsilon$, the estimator breaks down due to the finite number of samples used (finite size effect) after the network's giant component decomposes into smaller and smaller disconnected components. The Erd\H{o}s-R\'enyi approximation yields a critical percolation threshold $\varepsilon_c=1/(2(N-1))\approx 5 \cdot 10^{-4}$ for both maps using the parameters of Fig.~\ref{fig:one_dim_map_apl}, which is one order of magnitude smaller than the numerically observed phase transition point (Fig.~\ref{fig:one_dim_map_apl}). As explained in Sec.~\ref{sec:recurrence_threshold}, this is because the Erd\H{o}s-R\'enyi model does not account for the effects of spatial embedding and clustering. To be able to use the relationship of Eq.~(\ref{eq:critical_threshold}) based on the empirical results of~\cite{Dall2002} for $m=1$, we approximate $\rho(\varepsilon)=2\varepsilon$ (Bernoulli map) and $\rho(\varepsilon)=8\varepsilon\, \text{artanh}(1-2\varepsilon)/\pi^2$ (logistic map) for small $\varepsilon$. This yields $\varepsilon_c\approx 6.4 \cdot 10^{-3}$ for the Bernoulli map and $\varepsilon_c\approx 6.2 \cdot 10^{-3}$ for the logistic map, which is consistent with the phase transition points observed numerically (Fig.~\ref{fig:one_dim_map_apl}). The good agreement of predicted and observed phase transition for the Bernoulli map can be explained by the fact that the latter exactly meets the assumptions underlying the theory of Dall and Christensen (Sec. \ref{sec:recurrence_threshold}). These observations indicate that Eq.~(\ref{eq:critical_threshold}) is indeed useful to derive an educated guess for the proper choice of $\varepsilon$, even for strongly varying probability densities $p$. Moreover, the phase transition for the logistic map occurs at notably larger $\varepsilon$ than for the Bernoulli map. Consistently with the results of~\cite{Kong2007}, this indicates that the increased spatial clustering induced by peaks in the density $p$ leads to larger values of the critical mean degree $z_c$ and therefore the associated percolation threshold $\varepsilon_c$. However, the results of~\cite{Barnett2007} suggest that there may be in fact no true phase transition in giant component size for non-uniform $p$ in the limit $N \to \infty$.

For large $\varepsilon$, the approximation in Eq.~(\ref{eq:shortest_path_length}) is not valid anymore and, hence, the discrete estimator breaks down in this regime. Note that $\mathcal{L}(\varepsilon_{c'}) = 1$ for a critical $\varepsilon_{c'}$. Since $\mathcal{L}(\varepsilon) < 1$ whereas $\hat{\mathcal{L}}(\varepsilon)=1$ for $\varepsilon > \varepsilon_{c'}$, the definition of the discrete estimator is not meaningful anymore for thresholds larger than the critical threshold. For the Bernoulli map we have $\varepsilon_{c'}=1 / 3$ and for the logistic map, $\varepsilon_{c'}=4/\pi^2$ follows.

\begin{figure}[tbp]
   \centering
   \includegraphics[width=0.49\columnwidth]{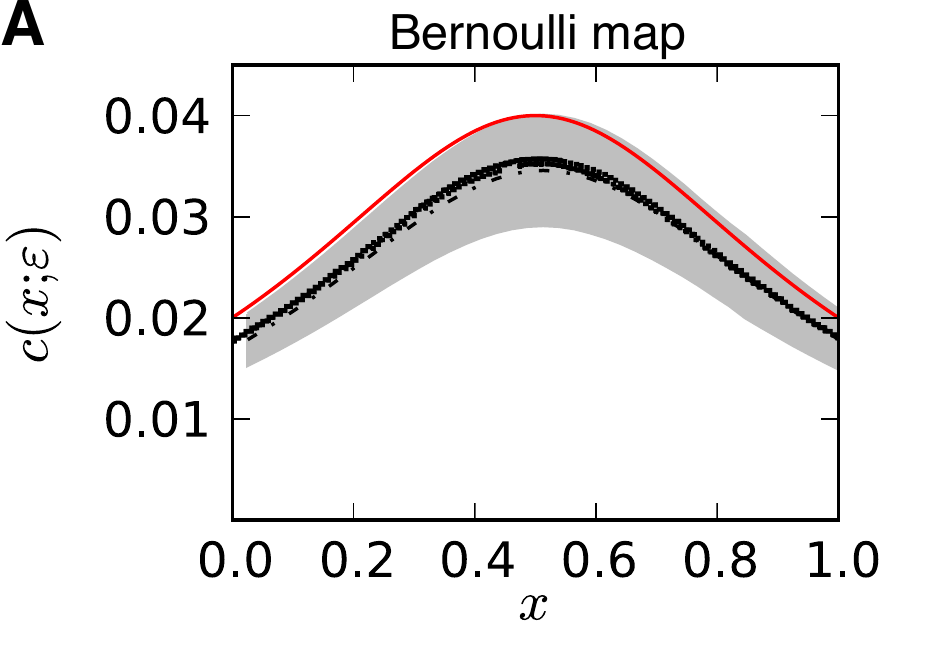}
   \includegraphics[width=0.49\columnwidth]{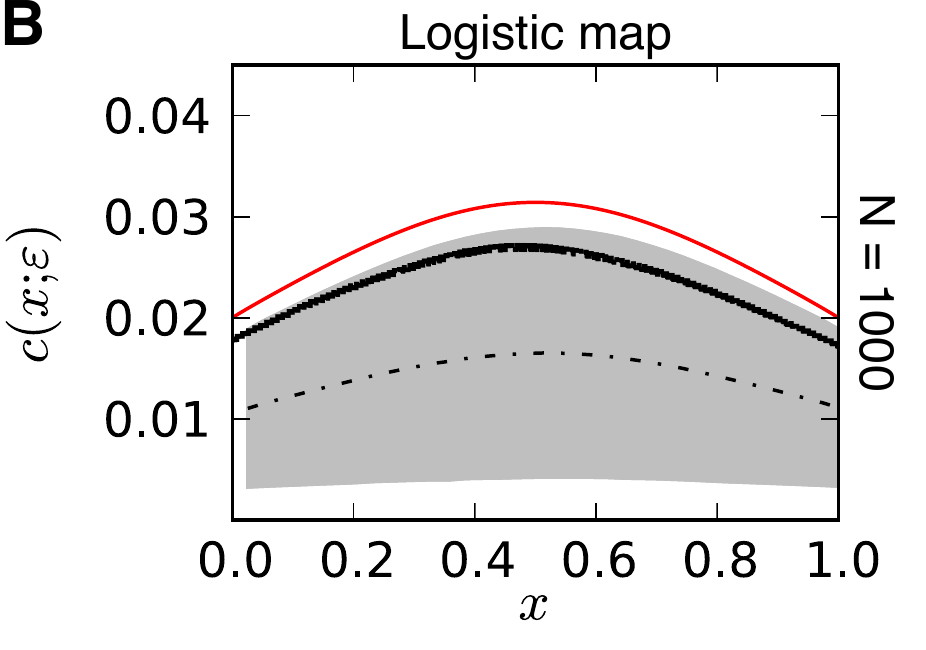}\\
   \includegraphics[width=0.49\columnwidth]{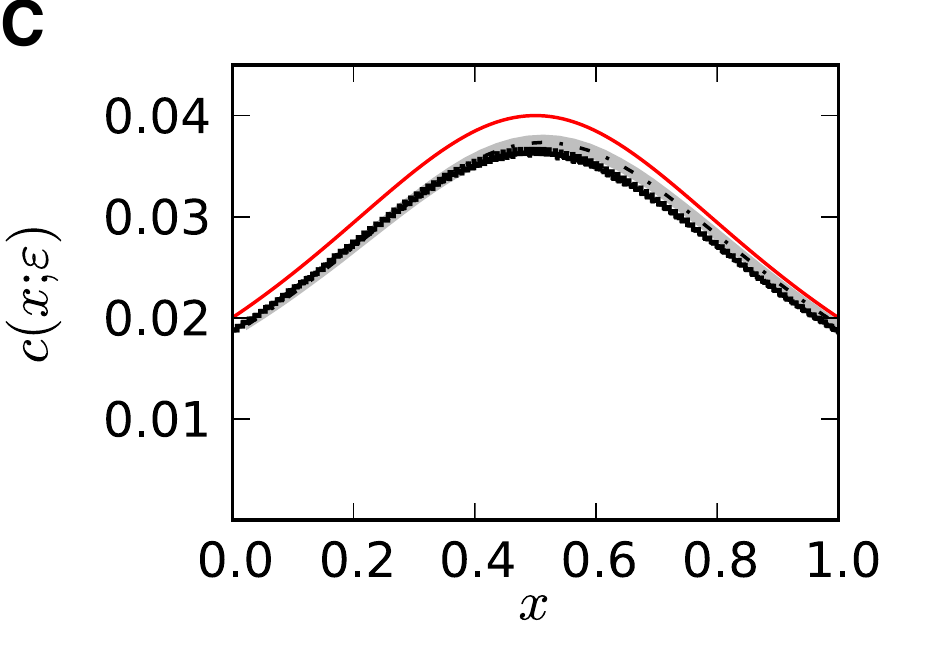}
   \includegraphics[width=0.49\columnwidth]{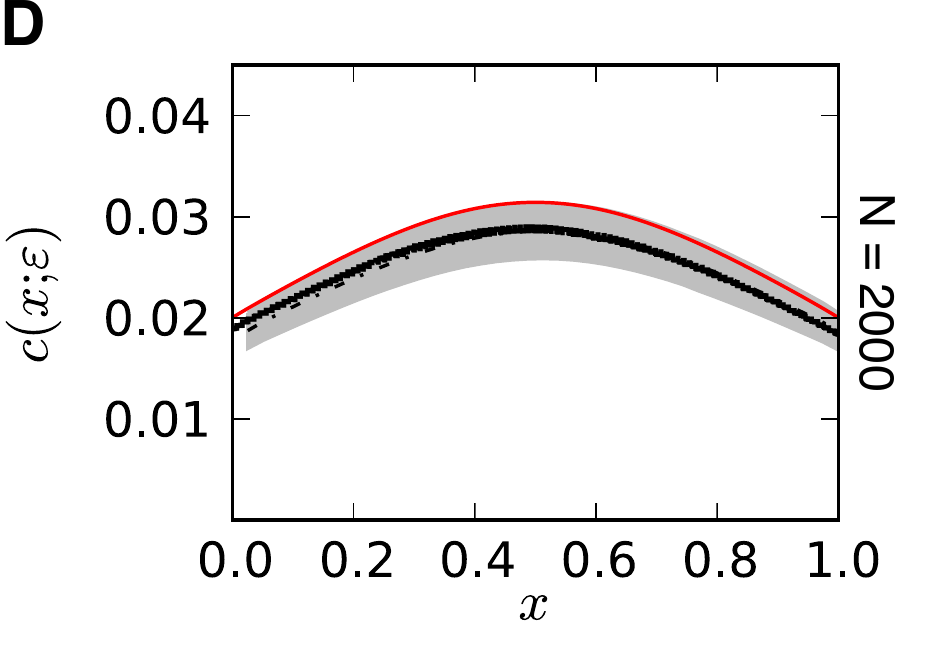}\\
   \includegraphics[width=0.49\columnwidth]{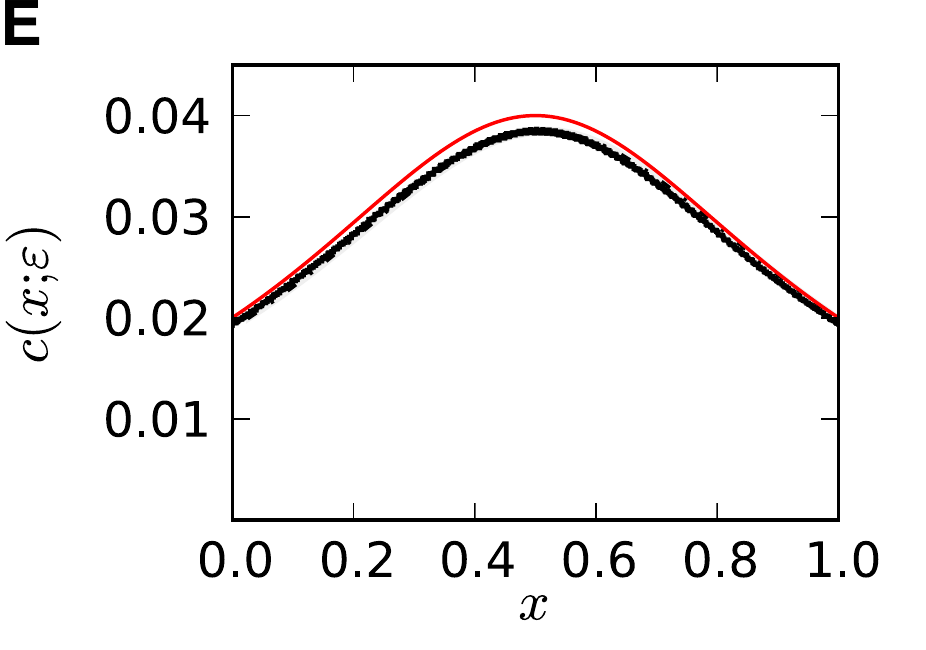}
   \includegraphics[width=0.49\columnwidth]{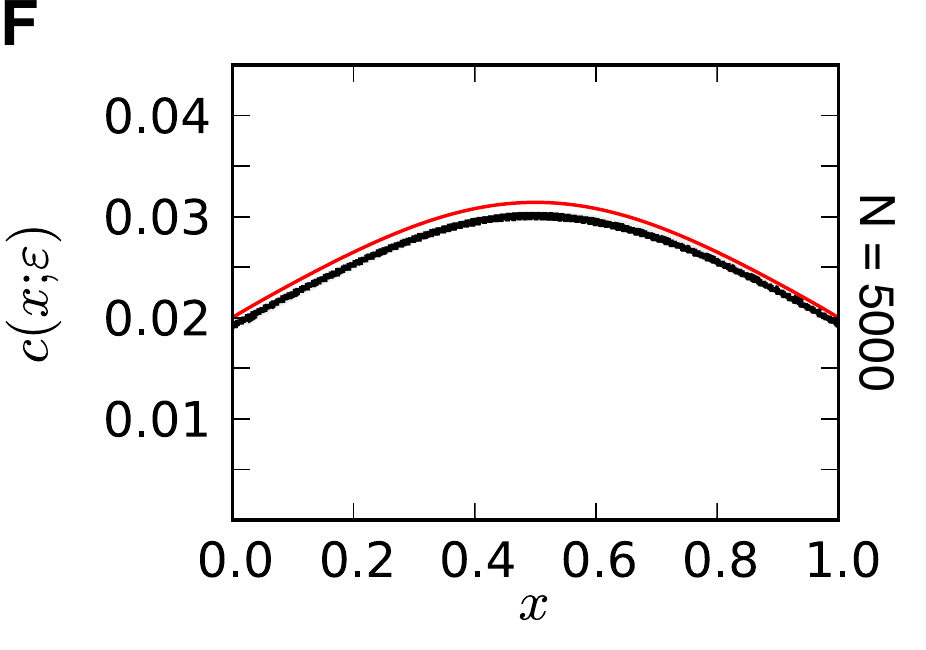}\\
   \caption{(Color online) Continuous $\varepsilon$-closeness $c(x;\varepsilon)$ for (A,C,E)~the Bernoulli map and (B,D,F)~the logistic map. Analytical results are indicated by red solid lines. Estimates $\hat{c}(x;\varepsilon)$ have been obtained from $\varepsilon$-recurrence networks at $\varepsilon=0.01$ constructed from single realizations of (A,B)~$N=1,000$, (C,D)~$N=2,000$ and (E,F)~$N=5,000$ samples (all: black squares). Ensemble mean (black dashed-dotted lines) and standard deviation (gray bands) have been calculated as in Fig. \ref{fig:one_dim_map_apl}. The standard deviation is too small to be visible in the plots for $N=5,000$~(E,F).}
   \label{fig:one_dim_map_closeness}
\end{figure}

The continuous $\varepsilon$-closeness $c(x;\varepsilon)$ is approximated well by the estimator $\hat{c}(x;\varepsilon)$ for both the Bernoulli and logistic maps (Fig.~\ref{fig:one_dim_map_closeness}). However, $\hat{c}(x;\varepsilon)$ is notably smaller than the true theoretical value particularly in the center of $S$ at $x=1/2$, implying that shortest paths are longer in the empirical $\varepsilon $-recurrence network than expected theoretically. This is clearly a finite size effect as the bias and variance of the estimator clearly decrease for growing $N$ and fixed $\varepsilon$ (Fig.~\ref{fig:one_dim_map_closeness}).

\begin{figure}[tbp]
   \centering
   \includegraphics[width=0.49\columnwidth]{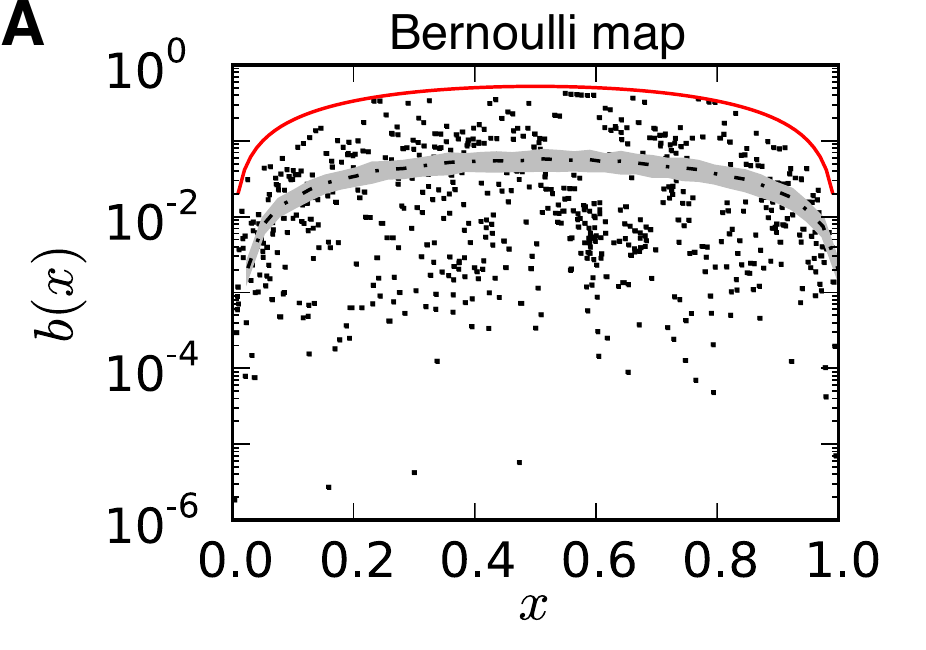}
   \includegraphics[width=0.49\columnwidth]{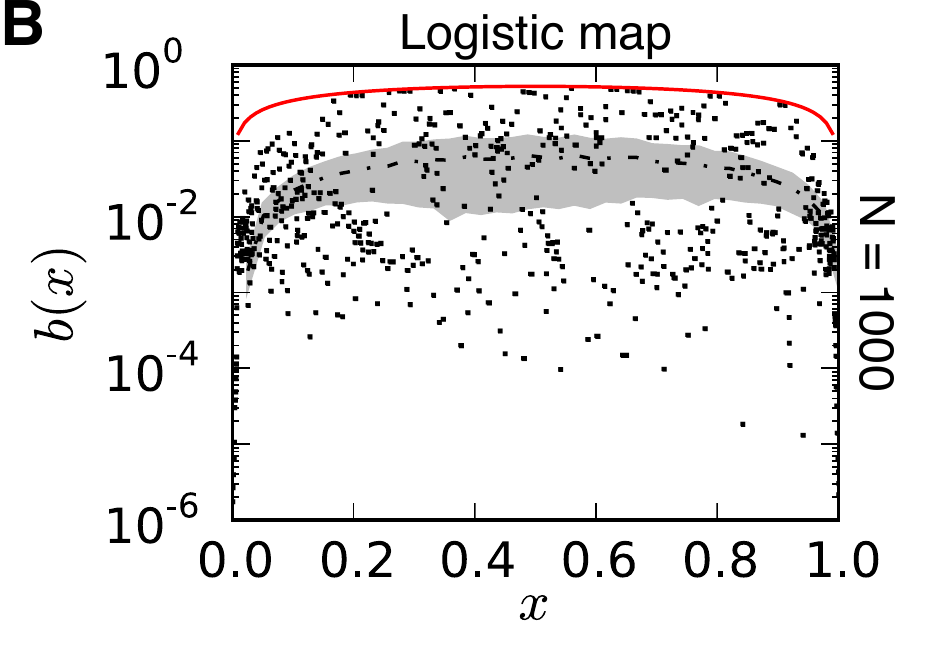}\\
   \includegraphics[width=0.49\columnwidth]{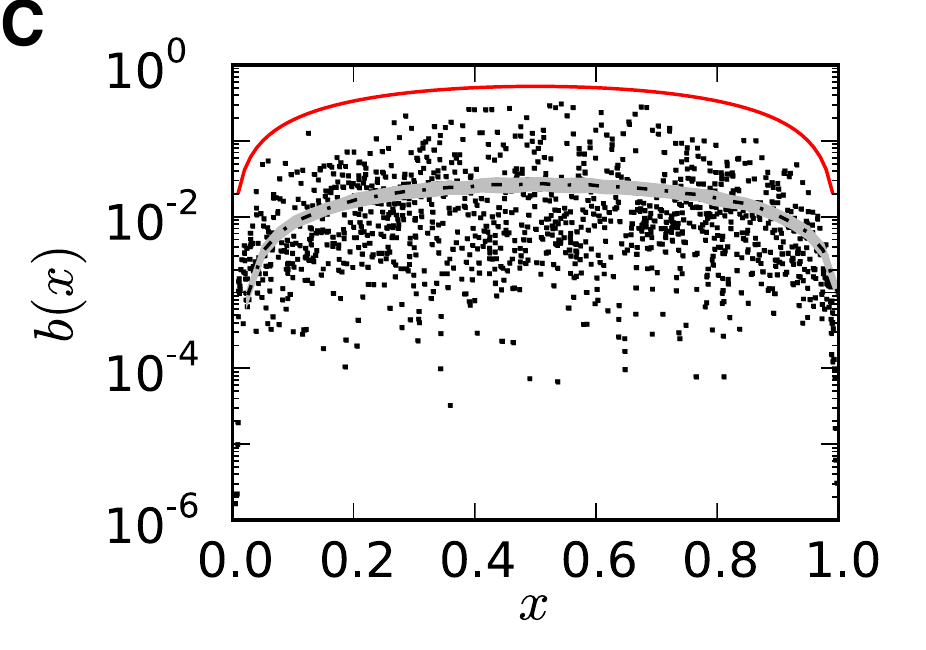}
   \includegraphics[width=0.49\columnwidth]{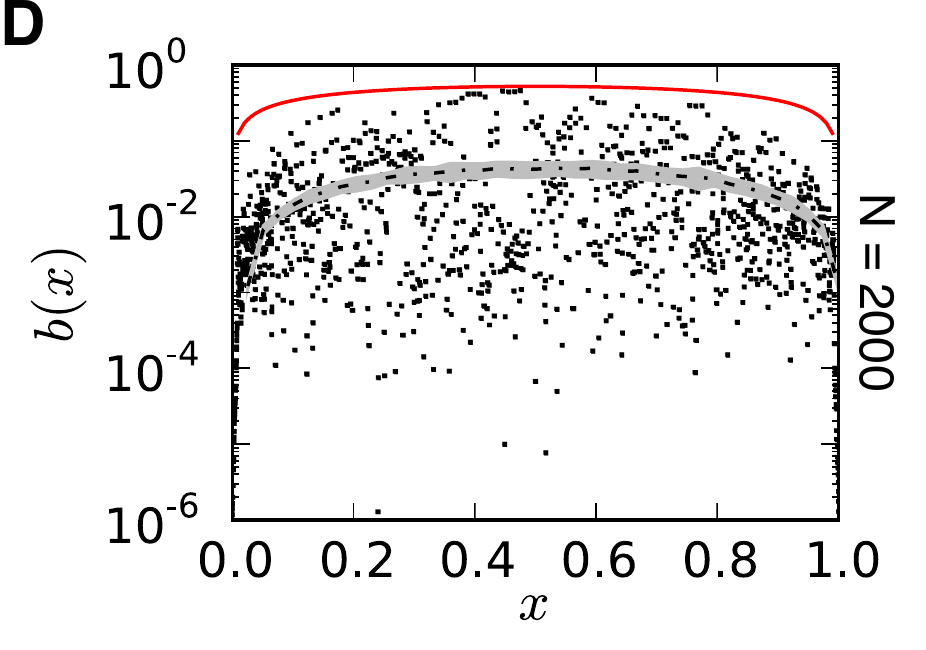}\\
   \includegraphics[width=0.49\columnwidth]{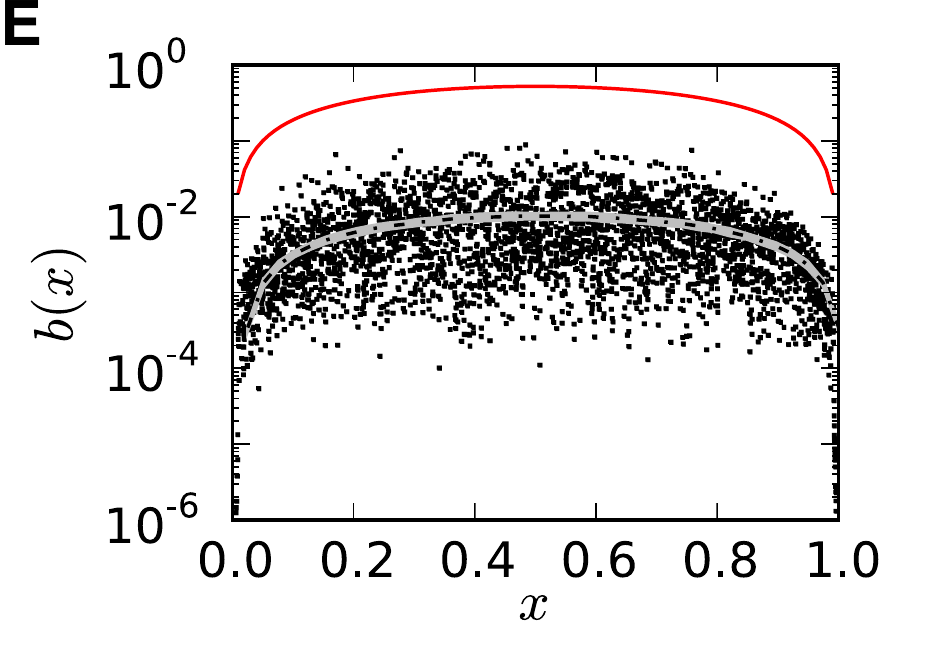}
   \includegraphics[width=0.49\columnwidth]{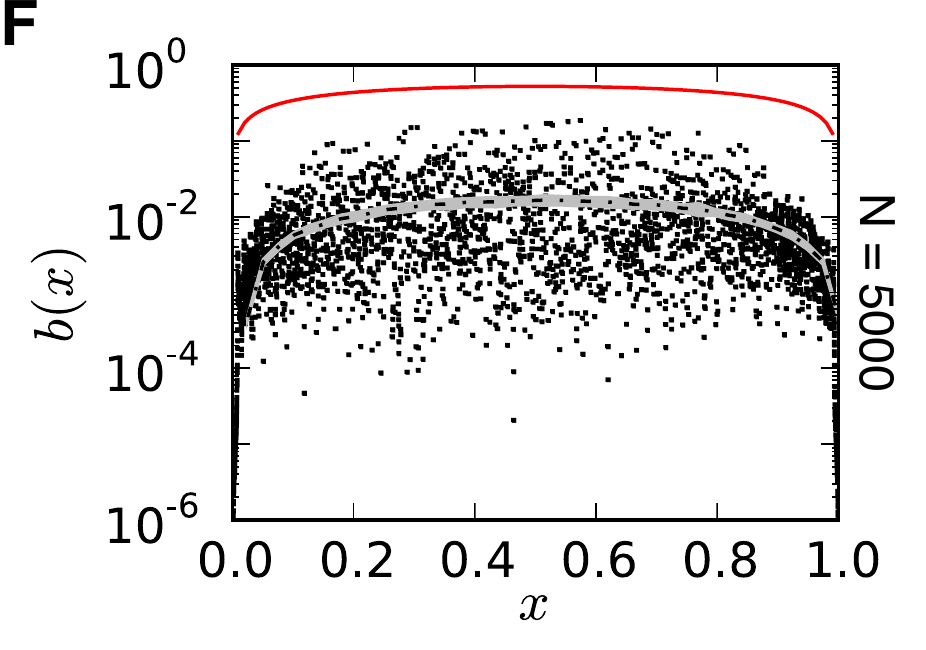}\\
   \includegraphics[width=0.49\columnwidth]{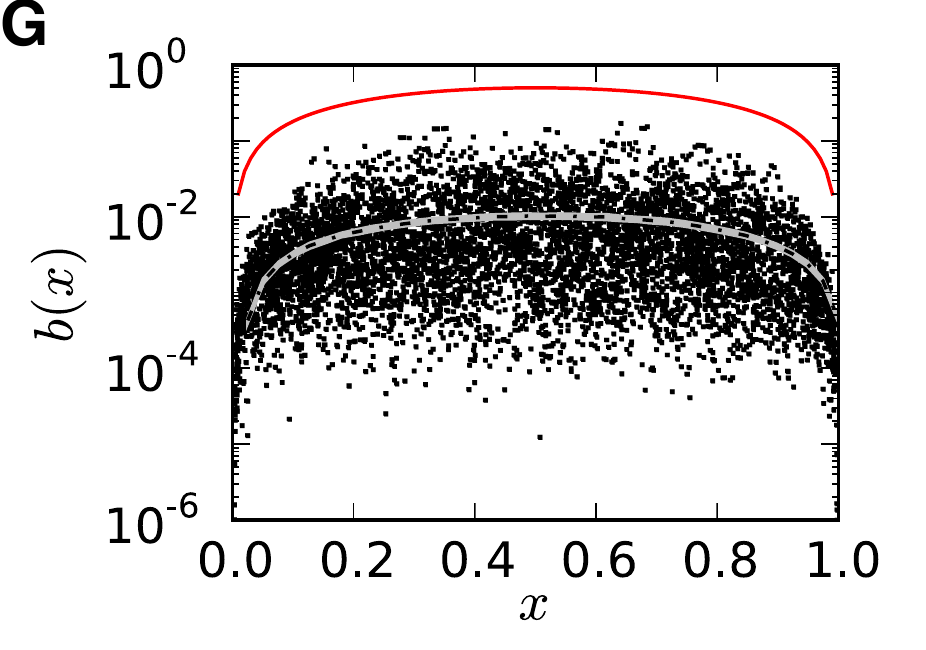}
   \includegraphics[width=0.49\columnwidth]{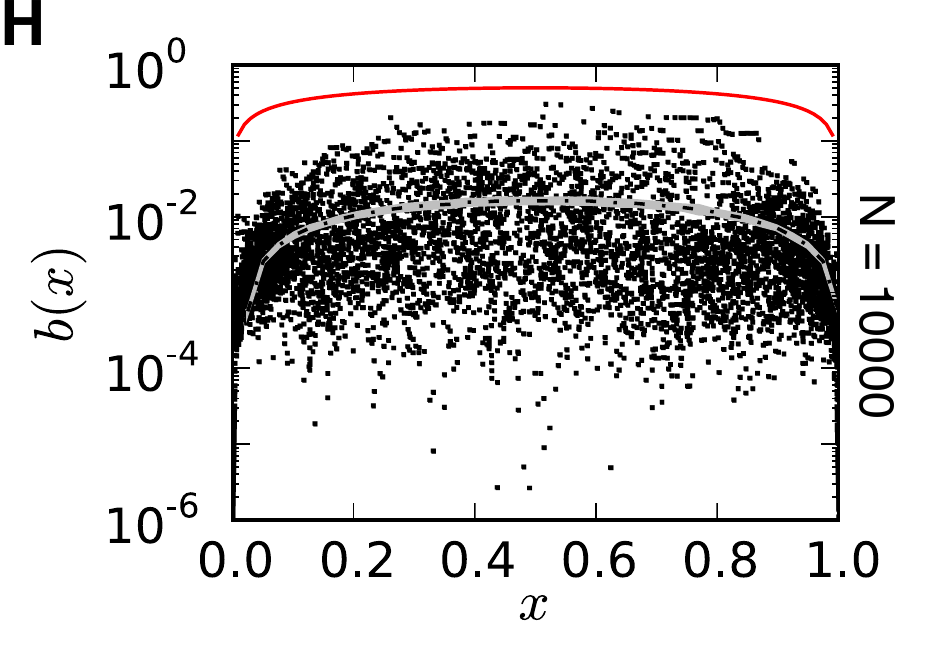}\\
   \caption{(Color online) Continuous $\varepsilon$-shortest path betweenness $b(x)$ for (A,C,E,G)~the Bernoulli map and (B,D,F,H)~the logistic map. Analytical results are indicated by red solid lines. Estimates $\hat{b}(x;\varepsilon)$ have been obtained from $\varepsilon$-recurrence networks at $\varepsilon=0.01$ constructed from single realizations of (A,B)~$N=1,000$, (C,D)~$N=2,000$ and (E,F)~$N=5,000$ samples (all: black squares). Panels~(G,H) show results for $\varepsilon=0.005$ and $N=10,000$. Ensemble mean (black dashed-dotted lines) and standard deviation (gray bands) have been calculated as in Fig. \ref{fig:one_dim_map_apl}. Note that $\hat{b}(x)$ is the discrete shortest path betweenness of~\cite{Donner2010b}, but normalized by its theoretical maximum value $(N-1)(N-2)/2$.}
   \label{fig:one_dim_map_betweenness}
\end{figure}

The shape of continuous $\varepsilon$-shortest path betweenness $b(x)$ is approached well by the estimator $\hat{b}(x;\varepsilon)$ for both maps (Fig.~\ref{fig:one_dim_map_betweenness}). However, there is a large bias that increases with the number of samples $N$, while the variance decreases with growing $N$. That the estimator $\hat{b}(x;\varepsilon)$ is generally smaller than the theoretical value $b(x)$ for all $x$ can be explained by the skipping of vertices due to the finite $\varepsilon$ in the empirical $\varepsilon$-recurrence network. This effect is expected to increase for growing $N$ when $\varepsilon$ is fixed, since more and more vertices can be skipped along a shortest path for the same recurrence radius $\varepsilon$, which also explains the growing bias in this setting. Accordingly, the bias decreases for decreasing $\varepsilon$ when $N$ is sufficiently large regarding the discussion of suitable choices of $\varepsilon$ in Sec.~\ref{sec:recurrence_threshold} (Fig.~\ref{fig:one_dim_map_betweenness}G,H). However, the bias is not a problem in practical situations, because for local measures we are usually only interested in relative differences between vertices and not in the absolute values. 

\subsection{Periodic and quasi-periodic dynamics}

\subsubsection{Periodic orbit}

We analyze next a periodic orbit (general closed curve) of curve length $l$ embedded in an $m$-dimensional phase space, i.e., $S=\left\{x \in \mathbb{R}^m : x=f(s); s\in[0,l]; f(0)=f(l)\right\}$, with uniform probability density $p(x)=1 / l$. The geodesic distance of two points $x(s),x(t)$ along the curve is then given by $g(x(s),x(t)) = |s-t|$. This yields
\begin{eqnarray*}
\mathcal{T}(\varepsilon)&=&\mathcal{C}(\varepsilon) = \mathcal{C}(x;\varepsilon) = \frac{3}{4} \\
\mathcal{L}(\varepsilon) &=& \frac{l}{4} \varepsilon^{-1} \\
c(x;\varepsilon) &=& \frac{4 \varepsilon}{l} \\
b(x) &=& \frac{1}{4}.
\end{eqnarray*}
As the periodic orbit is a one-dimensional set we have $\mathcal{E}(\varepsilon)=0$ and $e(x;\varepsilon)=\infty$ as above. For example, a circular orbit of radius $R$ as generated by a harmonic oscillator with $S=\left\{x \in \mathbb{R}^2 : x_1=R\sin(s / R), x_2=R\cos(s / R); s \in [0,2\pi R]\right\}$ and $p(x)=1 / (2 \pi R)$ gives the above results with $l=2\pi R$.

\subsubsection{Flat 2-torus}

Quasi-periodic dynamics is displayed by a system oscillating with two incommensurable frequencies $\omega_1$ and $\omega_2$, i.e., where the ratio $\omega_1/\omega_2$ is not a rational number. The phase space trajectory fills a 2-torus $S=\{x=(s,t): s \in [0, 2 \pi R] ,t \in [0,2 \pi r] \}$ uniformly with $p(s,t)=p=1 / (4 \pi^2 R r)$. The radii $R,r$ are related to the oscillation's amplitudes. With the geodesic distance
\begin{eqnarray*}
g((s,t),(s',t')) &=& \big( \min\left(|s-s'|, 2 \pi R - |s-s'| \right)^2 \\ && \quad + \min\left(|t-t'|, 2 \pi r - |t-t'| \right)^2 \big)^{\frac{1}{2}}
\end{eqnarray*}
we obtain
\begin{eqnarray*}
\mathcal{T}(\varepsilon)&=&\mathcal{C}(\varepsilon) = \mathcal{C}(x;\varepsilon) = 1 - \frac{3\sqrt{3}}{4\pi} \approx 0.5865 \\
\mathcal{L}(\varepsilon) &=& \frac{\pi \varepsilon^{-1}}{12 r R } \bigg(4 r R \sqrt{r^2+R^2}+3 r^3 \text{arsinh}\left(\frac{R}{r}\right) \\ && \quad+ 2 R^3 \text{artanh}\left(\frac{r}{\sqrt{r^2+R^2}}\right) \\ && \quad - r^3 \text{artanh}\left(\frac{R}{\sqrt{r^2+R^2}}\right)\bigg) \\
\mathcal{E}(\varepsilon) &=& 2 \pi  r R \varepsilon^{-1}  \bigg(2 r \text{artanh}\left(\frac{R}{\sqrt{r^2+R^2}}\right) \\ && \quad + R \, \text{ln}\left(\frac{r+\sqrt{r^2+R^2}}{-r+\sqrt{r^2+R^2}}\right)\bigg)^{-1}.
\end{eqnarray*}
Because of symmetry, the local path-based measures do not depend on $x$ as for the periodic orbit discussed above and we have
\begin{eqnarray*}
c(x;\varepsilon) &=& \mathcal{L}(\varepsilon)^{-1}\\
e(x;\varepsilon) &=& \mathcal{E}(\varepsilon)^{-1}\\
b(x) &=& \frac{1}{4 \pi^2 R r} = p.
\end{eqnarray*}
As expected, in the limit $r\to 0$ the average path length converges to the value obtained for a circle of radius $R$ (see above), i.e., $\lim_{r\to 0} \mathcal{L}(\varepsilon) = (\pi R/2) \varepsilon^{-1}$.

\subsection{Higher-dimensional symmetric sets}

The $m$-dimensional hyperball and hypercube may be viewed as tractable idealizations of higher dimensional attracting sets of dynamical systems (in this section, we set $d=m$, since the manifold dimension of $S$ is integer). Their study highlights that continuous path-based measures may depend sensitively and non-trivially on the global geometry of the set. In contrast, their neighborhood and transitivity-based counterparts just depend on the dimension $m$ and are therefore identical for the hyperball and hypercube~(Eq.~(\ref{eq:hypergeom_transitivity})). The sets considered here are convex, hence, $g(x,y)=\|x-y\|_2$ holds when using the Euclidean norm.

\subsubsection{$m$-dimensional hyperball}

\begin{figure}[tbp]
   \centering
   \includegraphics{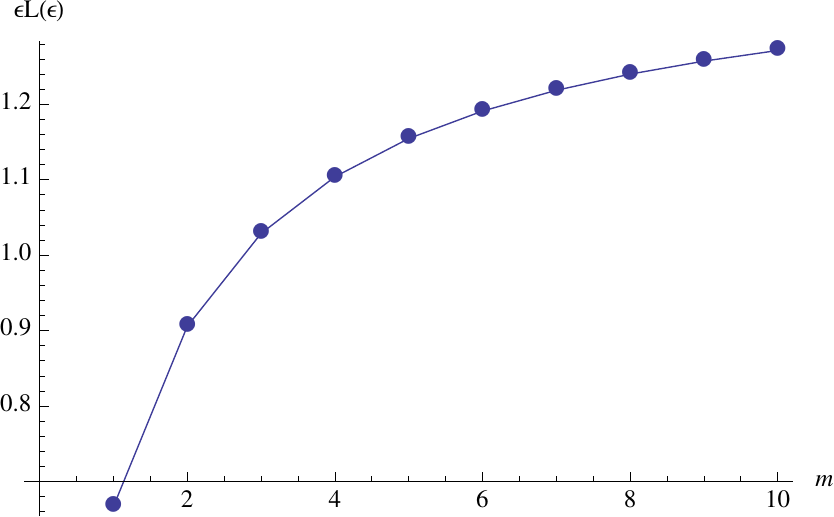} 
   \caption{Continuous $\varepsilon$-average path length $\varepsilon\mathcal{L}(\varepsilon)$ of the hyperball $\mathbb{S}^m$ with uniform probability density, obtained from Eq.~(\ref{eq:apl_hyperball}).}
   \label{fig:hyperball_apl}
\end{figure}

Here we consider the m-dimensional hyperball $S=\mathbb{S}^m$ with the uniform probability density $p(x)=p=1/\int_Sdx = \textrm{Vol}\left(\mathbb{S}^m\right)^{-1}=\Gamma\left(\frac{m}{2} +1\right) / \pi^\frac{m}{2}$. Following Hammersley~\cite{Hammersley1950}, the $r$-th moment of the distribution of point-to-point distances $\|x-y\|_2$ in $\mathbb{S}^m$ is given by
\begin{equation}
\mu_{mr} = 2 \: \frac{m\Gamma\left(m+1\right)}{\Gamma\left(\frac{1}{2}m+\frac{1}{2}\right)} \frac{\Gamma\left(\frac{1}{2}m + \frac{1}{2}r + \frac{1}{2}\right)}{(m+r)\Gamma\left( m+\frac{1}{2}r +1 \right)}. \label{eq:moments_m_sphere}
\end{equation}
Then the continuous $\varepsilon$-average path length is 
\begin{eqnarray}
\mathcal{L}(\varepsilon) &=& \mu_{m1} \varepsilon^{-1} \nonumber \\
&=& 2 \: \frac{m}{m+1} \frac{\Gamma\left(m+1\right)}{\Gamma\left(\frac{1}{2}m+\frac{1}{2}\right)} \frac{\Gamma\left(\frac{1}{2}m + 1\right)}{\Gamma\left( m+\frac{3}{2} \right)} \varepsilon^{-1}, \label{eq:apl_hyperball}
\end{eqnarray}
and all its higher moments are known via Eq.~(\ref{eq:moments_m_sphere}) (see Fig.~\ref{fig:hyperball_apl}). Some examples for lower-dimensional spheres follow:
\begin{equation*}
\varepsilon \mathcal{L}(\varepsilon) = \begin{cases} \frac{2}{3} & m=1 \\ \frac{128}{45} \pi^{-1} & m=2 \\ \frac{36}{35} & m=3 \\ \frac{16384}{4725}\pi^{-1} & m=4 \\ \frac{800}{693} & m=5 \end{cases}.
\end{equation*}

\noindent Note that the result for $m=1$ agrees with the corresponding one for the Bernoulli map when considering the stretching of the domain by a factor of $2$, since $\mathbb{S}^1=[-1,1]$. In the limit $m \to \infty$ the continuous $\varepsilon$-average path length is $\mathcal{L}(\varepsilon) = \sqrt{2} \varepsilon^{-1}$ (see Fig.~\ref{fig:hyperball_apl}). We can also derive in closed form an expression for the continuous $\varepsilon$-closeness $c(0;\varepsilon)$ of the center of $\mathbb{S}^m$, taking advantage of the spherical symmetry:
\begin{eqnarray*}
c(0;\varepsilon)^{-1} &=& \varepsilon^{-1} \int_{\mathbb{S}^m}dx_1 \dots dx_m \: p \sqrt{x_1^2  + \dots + x_m^2}\\
&=& \varepsilon^{-1} \Omega_m p \int_0^1 dr \: r^{m-1} r.
\end{eqnarray*}
With the full solid angle in $m$ dimensions $\Omega_m=m\pi^\frac{m}{2} / \Gamma(\frac{m}{2} + 1)$ this leads to
\begin{equation}
c(0;\varepsilon) = \frac{m+1}{m} \varepsilon.
\end{equation}
Note that the limit $\varepsilon c(0;\varepsilon)^{-1} \to 1$ for $m \to \infty$ shows that almost all of the measure $\mu(\mathbb{S}^m)$ of the unit radius hyperball $\mathbb{S}^m$ is concentrated at its surface for large $m$. For the special case of $m=2$ (unit disk with uniform $p(x)$), Lew \textit{et al.}~\cite{Lew1978} give a nearly closed-form expression for the continuous $\varepsilon$-closeness at $x(q)=(q,0)$,
\begin{equation*}
c(x(q);\varepsilon)^{-1} = \frac{1}{9\pi} \left( 16(q^2-1)K(q^2) + 4(q^2+7)E(q^2) \right) \varepsilon^{-1},
\end{equation*}
where $0\leq q\leq 1$ and the value for arbitrary $x\in\mathbb{S}^2$ may be obtained after an appropriate rotation. $K(m)$ and $E(m)$ are complete elliptic integrals of the first and second kind (see \S17.3 in~\cite{Abramowitz1964}).

For the continuous local $\varepsilon$-efficiency $e(0;\varepsilon)$ of the center of $\mathbb{S}^m$ we get for $m>1$
\begin{eqnarray*}
e(0;\varepsilon) &=& \varepsilon \int_{\mathbb{S}^m}dx_1 \dots dx_m \: p \sqrt{x_1^2 + \dots + x_m^2}^{-1}\\
&=& \varepsilon \Omega_m p \int_0^1 dr \: r^{m-1} r^{-1}\\
&=& \frac{m}{m-1} \varepsilon.
\end{eqnarray*}
A somewhat more involved calculation of the continuous $\varepsilon$-betweenness $b(0)$ of the center of $\mathbb{S}^m$ yields (see Appendix~\ref{appx:betweenness_hyperball})
\begin{equation}
b(0) = \frac{1}{\Omega_m} = \frac{\Gamma(\frac{m}{2} + 1)}{m \pi^{\frac{m}{2}}}.
\end{equation}
The high degree of symmetry of $\mathbb{S}^m$ allows us to derive closed-form results for local path-based measures at its center for many $p(x)=p(r,\Omega)$, as long as the probability density separates into a radial and an angular part, i.e., $p(r,\Omega)=p(r)p(\Omega)$.

\subsubsection{$m$-dimensional hypercube}

The hypercube $S=\mathbb{K}^m=[0,1]^m$ with uniform probability density $p(x)=\textrm{Vol}(\mathbb{K}^m)^{-1}=1$ is much harder to treat analytically than the hyperball $\mathbb{S}^m$. Hence, rigorous results are only available for isolated dimensions $m$ and a subset of the continuous measures defined above~\cite{Anderssen1976,Bailey2006,Bailey2007,Bailey2010}. Solving the resulting general box integrals remains a largely unsolved problem of applied and experimental mathematics. 

\begin{figure}[tbp]
   \centering
   \includegraphics{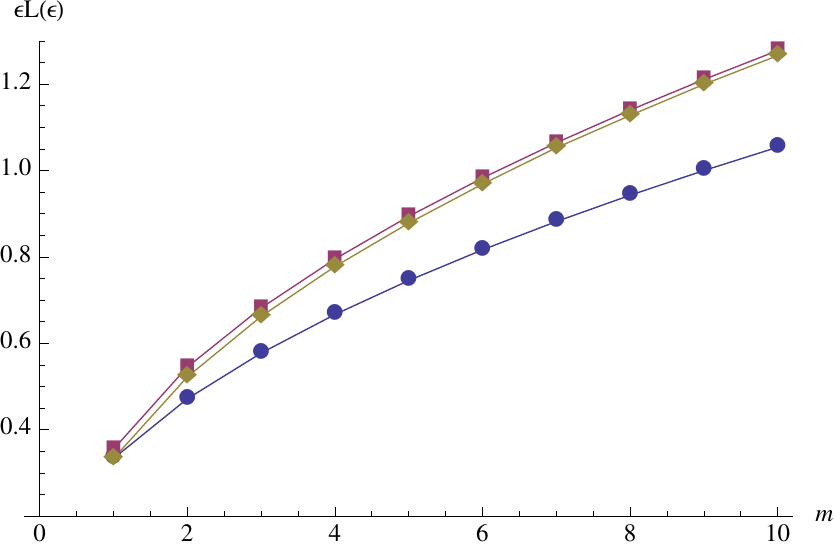} 
   \caption{(Color online) Continuous $\varepsilon$-average path length $\varepsilon\mathcal{L}(\varepsilon)$ of the hypercube $\mathbb{K}^m$ with uniform probability density, obtained by numerical Monte Carlo integration using Mathematica (yellow diamonds). Analytical lower (blue discs) and upper (red squares) bounds from~\cite{Anderssen1976} are also shown.}
   \label{fig:hypercube_apl}
\end{figure}

The following closed-form expressions for the continuous $\varepsilon$-average path length $\mathcal{L}(\varepsilon)$ are based on the expectation values for point-to-point distances $\Delta_{\mathbb{K}^m}(1)=\varepsilon\mathcal{L}(\varepsilon)$ (see Eq.~(\ref{eq:delta_def})) listed in~\cite{Bailey2010}:
\begin{equation*}
\varepsilon\mathcal{L}(\varepsilon) = \begin{cases}
\frac{1}{3} & m=1 \\
\frac{1}{15} \left(2+\sqrt{2}+5 \ln\left(1 + \sqrt{2}\right)\right) & m=2 \\
-\frac{118}{21} - \frac{2}{3}\pi + \frac{34}{21}\sqrt{2} - \frac{4}{7}\sqrt{3} & m=3 \\ \quad + 2 \ln(1+\sqrt{2}) + 8 \ln(\frac{1+\sqrt{3}}{\sqrt{2}}) & \\
\end{cases}.
\end{equation*}
Some numerical results for $m=1,\dots,10$ are displayed in Fig.~\ref{fig:hypercube_apl}. Anderssen \textit{et al.}~\cite{Anderssen1976} proved the bounds 
\begin{equation}
\frac{1}{3}\sqrt{m} \leq \varepsilon \mathcal{L} \leq \sqrt{\frac{1}{6} m} \sqrt{\frac{1}{3} \left(1 + 2\sqrt{1 - \frac{3}{5m}}\right)}
\end{equation}
implying $\varepsilon \mathcal{L} \to \infty$ for $m\to\infty$. This is in contrast to the hyperball, where this limit is finite (see above). Using expectation values for the inverse point-to-point distances $\Delta_{\mathbb{K}^m}(-1)= \varepsilon^{-1}\mathcal{E}(\varepsilon)^{-1}$, we are able to give the following expressions for the continuous $\varepsilon$-efficiency $\mathcal{E}(\varepsilon)$:
\begin{equation*}
(\varepsilon\mathcal{E}(\varepsilon))^{-1} = \begin{cases}
\infty & m=1 \\
\frac{4}{3}(1-\sqrt{2}) + 4\ln(1+\sqrt{2}) & m=2\\
\frac{2}{5} - \frac{2}{3} \pi + \frac{2}{5} \sqrt{2} - \frac{4}{5} \sqrt{3} & m=3 \\ \quad + 2\ln(1+\sqrt{2}) \\ \quad + 12 \ln\left(\frac{1+\sqrt{3}}{\sqrt{2}}\right)  \\ \quad - 4 \ln(2+\sqrt{3}) & \\
\end{cases}.
\end{equation*}
Note that as $S=\mathbb{K}^1=[0,1]$, the results for $m=1$ agree with the corresponding ones for the Bernoulli map for both continuous $\varepsilon$-average path length and efficiency. Further expressions for $\Delta_{\mathbb{K}^m}(1)$ and $\Delta_{\mathbb{K}^m}(-1)$ for $m=4,5$ are given in~\cite{Bailey2010}.

Another object of interest in the theory of box integrals is the integral
\begin{equation}
\tilde{B}_m(\eta) = \int_{\mathbb{K}^m} dx \|x\|_2^\eta,
\end{equation}
which is related to the continuous $\varepsilon$-closeness $c(0;\varepsilon)$ of the origin $x=0$ (and, by symmetry, to that of all the $2^m$ corners of the hypercube) for $\eta=1$ and to the local efficiency $e(0;\varepsilon)$ of the same points for $\eta=-1$:
\begin{eqnarray}
c(0;\varepsilon) &=& \tilde{B}_m(1)^{-1} \varepsilon \\
e(0;\varepsilon) &=& \tilde{B}_m(-1) \varepsilon.
\end{eqnarray}
We can now once again use results from Bailey \textit{et al.}~\cite{Bailey2010} to give some closed forms for small $m$:
\begin{equation*}
\varepsilon c(0;\varepsilon)^{-1} = \begin{cases}
\frac{1}{2} & m=1 \\
\frac{1}{3}(\sqrt{2} + \ln(1+\sqrt{2})) & m=2 \\
\frac{1}{4}\sqrt{3} - \frac{1}{24} \pi + \frac{1}{2}\ln(2 + \sqrt{3}) & m=3
\end{cases}
\end{equation*}
and
\begin{equation*}
\varepsilon^{-1} e(0;\varepsilon) = \begin{cases}
\infty & m=1 \\
2\ln(1+\sqrt{2}) & m=2 \\
-\frac{1}{4}\pi + \frac{2}{3}\ln(2+\sqrt{3}) & m=3
\end{cases}.
\end{equation*}
Further solutions for $m=4,5$ are given in~\cite{Bailey2010}.

\section{Discussion}
\label{sec:discussion}

We have shown that the definitions of continuous geometric measures provided in this paper are feasible for describing $\varepsilon$-recur\-rence networks for time series analysis as well as, more generally, random geometric graphs~\cite{Dall2002,Herrmann2003,Penrose2003}. {Our theoretical framework may readily be generalized to encompass other classes of random networks with spatial constraints (i.e., spatial networks)~\cite{Barnett2007,Itzkovitz2005,Barthelemy2011}, e.g., those with an edge length distribution of the form $p_l(l) \propto \exp(-l/\xi)$ describing among others the substrate of climate networks~\cite{Tsonis2008a,Donges2009a} (in contrast to the sharp cutoff $p_l(l) \propto \Theta(\varepsilon - l)$ for random geometric graphs). For the neighborhood-based measures, this generalization can be achieved by substituting terms containing the Heaviside function or $B_{\varepsilon}(x)$ with suitably chosen expressions involving $p_l(l)$. One possible application to real-world spatial networks is computing expectation values for the characteristics of an ensemble of spatial random network surrogates to assess which properties of a given empirical network can be explained by $p_l(l)$ alone. Additionally, more general metrics could be used for measuring the distance $l$ between connected vertices.} Research along these lines may also help to shed light on the specific topology and dynamics of growing spatial complex networks (cf.~\cite{Kaiser2004,Zhang2006}).

Furthermore, we have demonstrated that the resulting continuous properties can be approximated by estimators calculated from empirical $\varepsilon$-recurrence networks reasonably well, even for relatively small $N$ and large $\varepsilon$. The continuous framework promotes considerable advances in the theoretical understanding of $\varepsilon$-recurrence-network-based time series analysis. Among others, from the examples of hyperballs and hypercubes in various dimensions $m$, the claim that path-based measures depend explicitly on the global geometry of the set $S$ is theoretically justified. This is in contrast to the continuous notions of local and global transitivity, as at least the continuous local $\varepsilon$-clustering coefficient $\mathcal{C}(x;\varepsilon)$ depends on the local dimensionality of the set $S$~\cite{Donner2010dimensions}. Along these lines, in the future we may gain an understanding of the differing performance of transitivity-based and path-based measures in classifying qualitatively different behavior of dynamical systems~\cite{Marwan2009,Zou2010}. For example, more complex dynamical systems such as the Lorenz and R\"ossler models or noisy dynamical systems, where no closed-form expression for the invariant probability density $p$ exists, may be studied by estimating $\hat{p}$ from simulated trajectories. $\varepsilon$-recurrence network measures could then be calculated by numerical integration techniques relying on $\hat{p}$ and the integral expressions given in this paper. Circumventing the computational limitations of discrete $\varepsilon$-recurrence network analysis when $N \to \infty$, this approach would in principle allow us to approximate the continuous geometric quantities defined above as closely as desired.

The examples of hyperballs and hypercubes establish links to some current research problems in probabilistic geometry and applied mathematics, among others, to the theory of box integrals~\cite{Anderssen1976,Bailey2006,Bailey2007,Bailey2010}. Perhaps these highly symmetric model sets could serve to understand theoretically some qualitative features of path-based $\varepsilon$-recurrence network measures for strange attractors such as the Lorenz or R\"ossler attractors. It remains an open question as to whether it is possible to solve the integrals for continuous path-based measures in the case of self-similar sets $S$ and more complex, potentially also self-similar densities $p$.

The theoretical framework put forward in this paper enables several practical advances, which are particularly relevant for applications to time series analysis of real-world data. Analytical solutions for continuous $\varepsilon$-recurrence network measures allow us to assess the bias and variance of the discrete estimators from complex network theory that have been used in the literature so far. These insights led to devising improved discrete estimators based on the concept of node-weighted network statistics~\cite{Heitzig2011}. Furthermore, for the first time we were able to formulate a theoretically motivated criterion for the selection of the recurrence threshold $\varepsilon$ based on the critical percolation threshold $\varepsilon_c$ which for a given system can be estimated using our theory.

Finally, we should note that we now have a comprehensive continuous theory for essentially all relevant measures of $\varepsilon$-recurrence networks. This foundation will help to further increase our understanding as well as strengthen the general confidence in the method of $\varepsilon$-recurrence network analysis in practical situations, e.g., the analysis of real-world time series. Our results suggest that $\varepsilon$-recurrence network analysis is the simplest and best understood network-based approach to nonlinear time series analysis available so far.

\appendix

\section{Sketch of proof of Eq.~(\ref{eq:shortest_path_length})}
\label{appx:proof_path_length}

For $\varepsilon>0$, we define the {\em continuous $\varepsilon$-distance} $l(x,y;\varepsilon)$ between $x\neq y\in S$ to be the smallest integer $k>0$ such that there are points $z_0,\dots,z_k\in S$ with $z_0=x$, $z_k=y$, and $||z_{i-1}-z_i||<\varepsilon$ for $i=1\ldots k$. Note that because $S$ is path-connected, $l(x,y;\varepsilon)$ is finite. We also put $l(x,x;\varepsilon)=0$. Let $l_{ij}(\varepsilon,N)\geq l(x_i,x_j;\varepsilon)$ be the network distance between $x_i$ and $x_j$ in the $\varepsilon$-recurrence network constructed from the first $N$ points of a sequence of independent draws from $p$. One can then prove that for fixed $\varepsilon$, fixed nodes $x_i,x_j$, and $N\to\infty$, it has probability one that $l_{ij}(\varepsilon,N)=l(x_i,x_j;\varepsilon)$ eventually (i.e., there is some $N(i,j,\varepsilon)$ so that $l_{ij}(\varepsilon,N)=l(x_i,x_j;\varepsilon)$ for all $N>N(i,j,\varepsilon)$). This is because for $k=l(x_i,x_j;\varepsilon)$, there is $\delta>0$ and $z_0,\dots,z_k\in S$ with $z_0=x$, $z_k=y$, and $||z_{i-1}-z_i||<\varepsilon-2\delta$ for $i=1\ldots k$, and with probability one, the sequence contains points $w_0,\dots,w_k\in S$ with $w_0=x_i$, $w_k=x_j$, and $||w_i-z_i||<\delta$ for $i=1\ldots k-1$, so that also $||w_{i-1}-w_i||<\varepsilon$ for $i=1\ldots k$, implying $l_{ij}(\varepsilon,N)\leq k$ when $N>N(i,j,\varepsilon)$ where $N(i,j,\varepsilon)$ is the index of the last of the $w_i$ to occur in the sequence. 

Moreover, $l(x,y;\varepsilon)\leq g(x,y)/\varepsilon+1$ and $\varepsilon l(x,y;\varepsilon)\leq g(x,y)$ if $g(x,y)$ is not an integer multiple of $\varepsilon$. This is because for all $\delta>0$ and $\varepsilon'<\varepsilon$, there is a path from $x$ to $y$ of length $\leq g(x,y)+\delta$, hence for $k=\lceil (g(x,y)+\delta)/\varepsilon'\rceil$ ($\lceil x \rceil$ is the smallest integer not less than x), there are $z_0,\dots,z_k\in S$ with $z_0=x$, $z_k=y$, and $||z_{i-1}-z_i||\leq\varepsilon'<\varepsilon$ for $i=1\ldots k$, so that $l(x,y;\varepsilon)\leq k$. 
On the other hand, if $S$ is sufficiently well-behaved, one will also have $\varepsilon l(x,y;\varepsilon)\nearrow g(x,y)$ for $\varepsilon\to 0$. More precisely, assume $S$ is ``locally almost convex'' in the sense that for all $L>1$, there is some $\varepsilon>0$ so that for all $x,y\in S$ with $||x-y||<\varepsilon$, we have $g(x,y)<L\varepsilon$. Then for all $L>1$, there is some $\varepsilon>0$ so that $\varepsilon l(x,y;\varepsilon)>g(x,y)/L$. Putting all these facts together, we see that $\varepsilon l_{ij}(\varepsilon,N)$ is a plausible estimate of $g(x,y)$.

\section{Continuous $\varepsilon$-average path length for Bernoulli map and uniformly distributed noise}
\label{appx:bernoulli_map}

For illustration, we give the detailed calculation of $\mathcal{L}(\varepsilon)$ for the Bernoulli map and, equivalently, uniformly distributed noise:
\begin{eqnarray*}
\varepsilon \mathcal{L}(\varepsilon) &=& \int_0^1 \int_0^1 dx dy \: |x-y| \\
&=& \int_0^1dx \: \left(\int_x^1 dy \: |x-y| + \int_0^x dy \: |x-y| \right) \\
&=& \int_0^1dx \: \left(\int_x^1 dy \: (y-x) + \int_0^x dy \: (x-y) \right) \\
&=& \int_0^1dx \: \left(\left[\frac{1}{2}y^2-xy\right]_x^1 + \left[xy-\frac{1}{2}y^2\right]_0^x \right) \\
&=& \int_0^1dx \: \left( \frac{1}{2}-x-\frac{1}{2}x^2+x^2 + x^2-\frac{1}{2}x^2 \right) \\
&=& \int_0^1dx \: \left( \frac{1}{2} -x + x^2 \right) \\
&=& \left[ \frac{1}{3}x^3 - \frac{1}{2}x^2 + \frac{1}{2}x \right]_0^1 = \frac{1}{3}.
\end{eqnarray*}

\section{Continuous $\varepsilon$-betweenness for the center of a hyperball}
\label{appx:betweenness_hyperball}

\begin{eqnarray*}
b(0) &=& p^2 \int\!\!\!\int_S dy\,dz \int_0^1dt\,\delta(f(t)) \\
&=& p^2 \int\!\!\!\int d\Omega\,d\Omega' \int_0^1\!\!\!\int_0^1 dr dr' r^{m-1} r'^{m-1} \delta(\Omega-\Omega') \\
&=& p^2 \int d\Omega \left(\int_0^1 dr\, r^{m-1} \right)^2\\
&=& p^2 \Omega_m \frac{1}{m^2} \\
&=& \frac{\Gamma\left(\frac{m}{2}+1\right)}{m\pi^{\frac{m}{2}}} = \frac{1}{\Omega_m}
\end{eqnarray*}

\begin{acknowledgements}
This work has been financially supported by the Leibniz association (project ECONS), the Federal Ministry for Education and Research (BMBF) via the Potsdam Research Cluster for Georisk Analysis, Environmental Change and Sustainability (PROGRESS), and IRTG~1740~(DFG). JFD thanks the German National Academic Foundation for financial support. We credit Roger Grzondziel and Ciaron Linstead for help with the IBM iDataPlex Cluster at the Potsdam Institute for Climate Impact Research. Complex network measures have been calculated using the software package \texttt{igraph}~\cite{Csardi2006}.
\end{acknowledgements}


\end{document}